\documentclass[11pt,letterpaper]{article}

\pdfoutput=1
\usepackage{jcappub}

\usepackage{appendix}
\usepackage{array}
\newcolumntype{x}[1]{>{\centering\arraybackslash}p{#1}}
\usepackage{bm}
\usepackage{color}
\usepackage{graphicx}
\usepackage{cancel}
\usepackage{amsfonts}
\usepackage{amssymb}
\usepackage{amsmath}
\usepackage{calc}
\usepackage{dcolumn}
\usepackage{mathrsfs}

\usepackage{soul}

\usepackage[normalem]{ulem}             % Underline and strike text etc.

\newcommand{\eg}{e.g.~}
\newcommand{\ie}{i.e.~}

\newcommand{\Eq}[1]{Eq.~\eqref{#1}}
\newcommand{\Fig}[1]{Fig.~\ref{#1}}

\newcommand{\Lag}{\mathscr{L}}	% Lagrangian
\newcommand{\beq}{\begin{equation}}
\newcommand{\eeq}{\end{equation}}

\newcommand{\ud}{\text{d}}
\newcommand{\bol}[1]{\boldsymbol{#1}}%\newcommand{\bol}[1]{\mathbf{#1}}

\newcommand{\ER}{E_\text{R}}
\newcommand{\Ed}{E'}

\newcommand{\vesc}{v_\text{esc}}
\newcommand{\vmin}{v_\text{min}}

\newcommand{\tmax}{t_\text{max}}
\newcommand{\tmin}{t_\text{min}}

\newcommand{\taumax}{\tau_\text{max}}
\newcommand{\taumin}{\tau_\text{min}}
\newcommand{\htaumin}{\hat{\tau}_\text{min}}

\newcommand{\ignore}[1]{}

\definecolor{rossoCP3}{cmyk}{0,.88,.77,.40}
\definecolor{verdeCP3}{rgb}{0.09765625, 0.57421875, 0.1015625}
\definecolor{bluCP3}{rgb}{0, 0.23, 0.67}

\ifx\pdfoutput\undefined
\usepackage[dvips,bookmarks]{hyperref}    % This is for arXiv.org
\else
\usepackage{hyperref}    % This is for pdftex
\fi
\hypersetup{colorlinks, bookmarksopen, bookmarksnumbered,
citecolor=verdeCP3, linkcolor=bluCP3, pdfstartview=FitH, urlcolor=rossoCP3}

\notoc
\newcommand{\AddrUCLA}{Department of Physics and Astronomy, UCLA, 475 Portola Plaza, Los Angeles, CA 90095 (USA)}
\newcommand{\AddrUNIPD}{Dipartimento di Fisica e Astronomia ``G.~Galilei'', Universit\`a di Padova and INFN, Sezione di Padova, Via Marzolo 8, 35131 Padova, Italy}

\begin{document}

\title{Prospects for detection of target-dependent annual modulation in direct dark matter searches}

\subheader{DFPD-2015/TH/28}

\author[a,b]{Eugenio Del Nobile,}
\author[a]{Graciela B.~Gelmini,}
\author[a]{Samuel J.~Witte}

\affiliation[a]{\AddrUCLA}
\affiliation[b]{\AddrUNIPD}

\emailAdd{delnobile@physics.ucla.edu}
\emailAdd{gelmini@physics.ucla.edu}
\emailAdd{switte@physics.ucla.edu}

\abstract{Earth's rotation about the Sun produces an annual modulation in the expected scattering rate at direct dark matter detection experiments. The annual modulation as a function of the recoil energy $\ER$ imparted by the dark matter particle to a target nucleus is expected to vary depending on the detector material. However, for most interactions a change of variables from $\ER$ to $\vmin$, the minimum speed a dark matter particle must have to impart a fixed $\ER$ to a target nucleus, produces an annual modulation independent of the target element. We recently showed that if the dark matter-nucleus cross section contains a non-factorizable target and dark matter velocity dependence, the annual modulation as a function of $\vmin$ can be target dependent. Here we examine more extensively the necessary conditions for target-dependent modulation, its observability in present-day experiments, and the extent to which putative signals could identify a dark matter-nucleus differential cross section with a non-factorizable dependence on the dark matter velocity.}

\keywords{dark matter theory, dark matter experiment}

\maketitle

\flushbottom

\newpage

%%%%%%%%%%%%%%%%%%%%%%%%%%%%%%%%%%%%%%%%%%%%%%%%%%%%%%%%%%%%%%%%%%%

\section{Introduction}

Despite being the dominant form of matter in the Universe, the exact nature of the dark matter (DM) is still unknown. One of the most well-motivated candidates for DM is a particle with few GeV to hundreds of TeV mass and weak-scale interactions, referred to as a weakly interacting massive particle (WIMP). Efforts to shed light on the non-gravitational interactions of WIMP DM primarily focus on either detecting the byproducts of DM annihilation or decay (indirect detection), producing DM in the laboratory through collisions of standard model particles, or detecting interactions between DM in the galactic halo and terrestrial nuclei (direct detection).

Direct DM detection experiments attempt to gain insight into both the particle physics properties of DM and the local DM velocity distribution by observing the energy deposited by DM particles interacting with nuclei as they pass through detectors. A key feature of any convincing direct detection signal would be the annual modulation of the scattering rate caused by Earth's rotation around the Sun~\cite{Drukier:1986tm}. For DM velocity distributions that are locally smooth and isotropic in the galactic frame, it is usually expected that the differential rate for dark matter scattering off a target nuclide $T$ is nearly sinusoidal and can be well represented by
\beq\label{modrate}
\frac{\ud R_T}{\ud \ER}(\ER, t) \simeq S_0(\ER) + S_\text{m}(\ER) \cos \! \left( \frac{2 \pi}{1 \text{ year}} (t - t_0) \right) ,
\eeq
where $\ER$ is the nuclear recoil energy. Allowing the modulation amplitude $S_\text{m}(\ER)$ to assume both positive and negative values, the phase $t_0$ is independent of $\ER$. Taking instead $S_\text{m}(\ER)$ to be non-negative, as we do in this paper, $t_0$ changes from early June at large $\ER$ to early December at small $\ER$, with the transition occurring sharply at a single $\ER$ value. Accounting for the presence of anisotropy in the DM halo modifies this picture, most notably by modifying the $\ER$ dependence of the modulation phase. The extent to which various forms of anisotropy, including DM substructure, the gravitational focusing (GF) of DM particles by the Sun, and triaxial halo models, modify~\Eq{modrate} has been investigated \eg in~\cite{Lee:2013wza,Bozorgnia:2014dqa,DelNobile:2015nua,Fornengo:2003fm,Green:2003yh,Green:2000jg,Savage:2006qr,Evans:2000gr,Gelmini:2000dm,Freese:2003na,Bruch:2008rx,Purcell:2012sh}.

At fixed recoil energies, experiments employing different target elements are not necessarily expected to measure the same modulation of the rate. However, for most interactions, some observables associated with the annual modulation like the modulation fraction or the time of maximum and minimum signal, $\tmax$ and $\tmin$, do not depend on the target nuclide when expressed as functions of $\vmin$. This is the minimum speed a DM particle must have in Earth's frame to impart a recoil energy $\ER$ on a target nucleus. This definition naturally treats $\vmin$ as an $\ER$-dependent function. Alternatively, it is possible to think of $\ER$ as a $\vmin$-dependent function. In this context, $\ER$ is interpreted as the extremum energy (corresponding to a maximum energy if the scattering is elastic, and either a maximum or minimum energy if the scattering is inelastic) that can be imparted to a nucleus by an incoming DM particle traveling with speed $v = \vmin$ in Earth's frame. For each nuclide there exists a bijective relation between $\ER$ and $\vmin$ dictated by the scattering kinematics, and the choice of one or the other as the independent variable may lead to different insights. As commented above, for most interactions (\eg the standard spin-independent (SI) and spin-dependent (SD) interactions) observables like $\tmax$ and $\tmin$ are nuclide-independent functions of $\vmin$ (this is no longer true when expressed as functions of $\ER$, since the $\ER$-$\vmin$ relation is target dependent). Therefore for studying the signal modulation for single-element targets it is convenient to adopt $\vmin$ as the independent variable (averaging over different isotopes). For targets consisting of multiple elements, one must choose whether to treat $\ER$ or $\vmin$ as the independent variable (see \eg\cite{Gondolo:2012rs, Gelmini:2015voa, DelNobile:2013cva}). When we consider multiple targets in Sec.~\ref{mdm} we choose to return to using $\ER$ as the independent variable.

We pointed out in~\cite{DelNobile:2015tza} that when the DM-nucleus differential cross section has a non-factorizable velocity dependence, as for DM interacting through a magnetic dipole or an anapole moment, $\tmax$ and $\tmin$ are no longer target-independent functions of $\vmin$. Here, we reconsider the analysis performed in~\cite{DelNobile:2015tza} and examine more extensively how target-dependent modulation arises, how various experiments can actually observe such a signal, and the extent to which putative signals could identify DM with a non-factorizable velocity dependence in its differential scattering cross section. Specifically, we consider how \textit{(i)} integrating the scattering rate over a finite energy range,  \textit{(ii)} the presence of multiple target elements with non-negligible contributions to the rate, and  \textit{(iii)} different DM-nucleus scattering kinematics affect the potential observability of target-dependent modulation.

This paper is organized as follows. In Section~\ref{DDintro} we introduce the formalism and discuss what conditions must be present for target-dependent modulation. In Section~\ref{MDMover} we take the particular example of DM interacting with nucleons through an anomalous magnetic dipole moment and discuss how observables associated with the annual modulation of the rate depend on $\vmin$ for specific targets employed in currents experiments. Additionally,  we examine how experiments would view a signal arising from magnetic dipole DM a function of the observed energy $\Ed$ and the extent to which the expected signal would be distinguishable from a signal arising from a standard SI or SD contact interaction, for both elastic and inelastic scattering. We conclude in Section~\ref{conclusion}.

\section{DM signal and its modulation}
\label{DDintro}

\subsection{Direct detection rate}
Direct DM detection experiments try to measure the recoil energy $\ER$ a nucleus initially at rest in the detector acquires after scattering with a DM particle with initial velocity $\bol{v}$ in the detector's rest frame. The differential scattering rate on a nuclide $T$ per unit detector mass is
\beq\label{diffrate}
\frac{\ud R_T}{\ud \ER}(\ER, t) = \frac{C_T}{m_T} \frac{\rho}{m} \int_{v \geqslant \vmin(\ER)} v \, f(\bol{v}, t) \, \frac{\ud \sigma_T}{\ud \ER} \, \ud^3 v \ ,
\eeq
where $\rho = 0.3$ GeV/cm$^3$ is the local DM density, $m$ is the DM particle mass, $C_T$ is the nuclide mass fraction in the detector, $m_T$ is the target nuclide mass, and $f(\bol{v}, t)$ is the DM velocity distribution in Earth's frame. The energy dependence of $\vmin(\ER)$, is dictated by the scattering kinematics, for instance for elastic scattering
\beq\label{vmin}
\vmin(\ER) = \sqrt{\frac{m_T \ER}{2 \mu_T^2}} \ .
\eeq

Experiments do not measure directly the recoil energy, but a proxy for it denoted here with $\Ed$. This detected energy can \eg be measured in keVee (keV electron-equivalent energy) or photoelectrons. For experiments that bin their data, the energy-integrated scattering rate between detected energies $\Ed_1$ and $\Ed_2$ is 
\beq
\label{rate}
R_{[\Ed_{1},\Ed_{2}]} (t) = \sum_T \int_{\Ed_1}^{\Ed_2} \ud \Ed \, \epsilon(\Ed) \int_{0}^{\infty} \ud \ER \, G_{T}(\ER, \Ed) \frac{\ud R_T}{\ud \ER} (\ER, t)  \ ,
\eeq
where $\epsilon(\Ed)$ is the counting efficiency and $G_{T}(\ER, \Ed)$ describes the probability that an event detected with energy $\Ed$ resulted from a nuclear recoil having energy $\ER$. $G_{T}(\ER, \Ed)$ is frequently taken to be a Gaussian with mean value $\langle \Ed \rangle = Q_T \ER$, where $Q_T(\ER)$ is an element-dependent quenching factor.

Typically one assumes the DM is on average at rest with respect to the galaxy, and the velocity distribution in the galactic frame $f_\text{G}(\bol{v})$ is smooth and isotropic. The DM velocity distribution in Earth's frame is then obtained via the Galilean transformation $f(\bol{v}, t) = f_\text{G}(\bol{v} + \bol{v}_\oplus(t) + \bol{v}_\odot)$, where $\bol{v}_\oplus(t)$ is the velocity of Earth with respect to the Sun and $\bol{v}_\odot$ the velocity of the Sun with respect to the galaxy. In this paper we choose to model the velocity of Earth with respect to the Sun following the procedure of Ref.~\cite{Lee:2013xxa}, and take the velocity of the Sun with respect to the Galaxy to be $\bol{v}_\odot=(11,232,7) $ km/s in galactic coordinates. Furthermore, for concrete applications we assume the Standard Halo Model (SHM), in which the velocity distribution of the dark halo is a truncated Maxwellian,
\beq
f_\text{G}(\bol{v}) = \frac{e^{- \bol{v}^2 / v_0^2}}{(\pi v_0^2)^{3/2} N_\text{esc}} \theta(\vesc - |\bol{v}|) \ ,
\eeq
with galactic escape velocity $\vesc = 533$ km/s~\cite{Piffl:2013mla} and velocity dispersion $v_0 = 220$ km/s~\cite{Kerr:1986hz}. The normalization,
\beq
N_\text{esc} = \text{Erf}(\vesc/v_0)-\frac{2\vesc}{\sqrt{\pi}v_0}e^{-\vesc^2/v_0^2} \ ,
\eeq
is chosen such that $\int \ud^3 v \, f_\text{G}(\bol{v}) = 1$.

DM that is on average at rest with respect to the Galaxy has a preferred direction of motion in the Sun's reference frame. For this reason, DM particles viewed in the Sun's reference frame appear as a constant ``wind", with velocities preferentially opposed to $\bol{v}_\odot$. The gravitational potential of the Sun bends the trajectories of DM particles as they pass by, resulting in a focusing effect that is maximized at Earth's location when Earth is on the leeward side with respect to the Sun, occurring on March $1^\text{st}$. This effect, referred to as GF, implies the DM density and velocity distribution at Earth's location depend on Earth's position relative to the Sun. The influence of GF is larger on slower moving particles as they spend more time in the Sun's gravitational potential, and is negligible on WIMPs traveling faster than a few hundred km/s in the Solar reference frame. The effect of GF is taken into account by replacing $f_\text{G}(\bol{v} + \bol{v}_\oplus(t) + \bol{v}_\odot)$ with $f_\text{G}(\bol{v}_\infty[\bol{v} + \bol{v}_\oplus(t)] + \bol{v}_\odot)$, where
\begin{equation}
\bol{v}_\infty[\bol{v}] = \frac{v_\infty^2\bol{v} + \frac{1}{2} v_\infty u_\text{esc}^2 \bol{\hat{r}} - v_\infty \bol{v}(\bol{v} \cdot \bol{\hat{r}})}{v_\infty^2 + \frac{1}{2} u_\text{esc}^2 - v_\infty(\bol{v} \cdot \bol{\hat{r}})}
\label{eq:vinfty}
\end{equation}
is the velocity a DM particle had asymptotically far away from the Sun's gravitational potential, such that its velocity when arriving at Earth is $\bol{v}$~\cite{Alenazi:2006wu}. Here $u_\text{esc} = \sqrt{2GM_\odot /r} \approx 40$ km/s is the escape velocity of the Solar System at Earth's location, $r$ is the Sun-Earth distance, $\bol{\hat{r}}$ is the unit vector pointing from the Sun to Earth, and $v_\infty^2 = v^2 - u_\text{esc}^2$.

\subsection{Time dependence of the rate}
For the commonly considered SI and SD contact interactions, the differential scattering cross section for DM-nucleus elastic scattering has the form
\beq\label{diffsigmastandard}
\frac{\ud \sigma_T}{\ud \ER}(\ER, v) = \frac{m_T \sigma_T F_T(\ER)^2}{2 \mu_T^2} \frac{1}{v^2} \ ,
\eeq
where $\mu_T$ is the DM-nucleus reduced mass, $\sigma_T$ is the total cross section for a point-like nucleus, and $F_T(\ER)$ is the appropriate nuclear form factor normalized as $F_T(0) = 1$. This general form arises every time the scattering amplitude for a point-like nucleus is (at least approximately) independent of the scattering angle, \ie of the recoil energy. In this case, 
\beq
\sigma_T \equiv \int_0^{\ER^\text{max}} \frac{\ud \sigma_T}{\ud \ER} \, \ud \ER = \ER^\text{max} \frac{\ud \sigma_T}{\ud \ER}
\eeq
where $\ER^\text{max} = 2 \mu_T^2 v^2 / m_T$ is the maximum recoil energy a nucleus can get from scattering elastically with a DM particle with speed $v$. The effect of the finite size of the nucleus is then taken into account with the appropriate form factor. The differential rate for cross sections of the form in \Eq{diffsigmastandard} then reads
\beq\label{diffratestandard}
\frac{\ud R_T}{\ud \ER}(\ER, t) = C_T \frac{\rho}{m} \frac{\sigma_T F_T(\ER)^2}{2 \mu_T^2} \, \eta_{0}(\vmin(\ER), t) \ ,
\eeq
with
\beq\label{standardeta}
\eta_{0}(\vmin, t) \equiv \int_{v \geqslant \vmin} \frac{f(\bol{v}, t)}{v} \, \ud^3 v \ .
\eeq
The modulation of the differential rate is determined solely by the time dependence in the velocity integral $\eta_{0}(\vmin, t)$, which is a target-independent function of $\vmin$, and therefore common to all experiments. Even though what enters the rate is the function $\eta_{0}(\vmin(\ER), t)$, which depends on the target through $\vmin(\ER)$, one can express $\ER$ as a function of $\vmin$ and study $\ud R_T / \ud \ER (\ER(\vmin), t)$, which is proportional to the target-independent quantity $\eta_{0}(\vmin, t)$ (see \eg\cite{DelNobile:2013cva, Gelmini:2015voa}).

The target-independent nature of the time dependence of the differential rate for the standard SI and SD contact interactions is a consequence of the fact that velocity and target dependence can be factored in the differential cross section shown in \Eq{diffsigmastandard}. One may then ask, in general, under what circumstances observables associated with the modulation of the rate, such as $\tmax$ and $\tmin$, are target-dependent functions of $\vmin$. Following our preliminary study~\cite{DelNobile:2015tza}, we find that this can only happen when the following conditions are met:
\begin{enumerate}
\item the velocity and target dependence in the differential cross section cannot be factored, and
\label{factorization}
\item the scattering events that can be recorded by an experiment probe portions of the DM velocity distribution that are locally anisotropic in the galactic frame.
\label{anisotropy}
\end{enumerate}

As shown in Ref~\cite{DelNobile:2015tza}, it is possible to meet both requirements and thus have a target-dependent modulation. Regarding point \ref{anisotropy}, anisotropy in the local DM velocity distribution can arise from an anisotropy in the smooth component of the halo, DM substructure, and gravitational interactions of DM with nearby massive objects such as the Sun. In this paper we choose to introduce anisotropy by only including the effect of GF of DM particles by the Sun because this anisotropy necessarily exists and is well understood~\cite{Alenazi:2006wu,Lee:2013wza,Bozorgnia:2014dqa}.

\begin{figure*}
\centering
\includegraphics[width=0.49\textwidth]{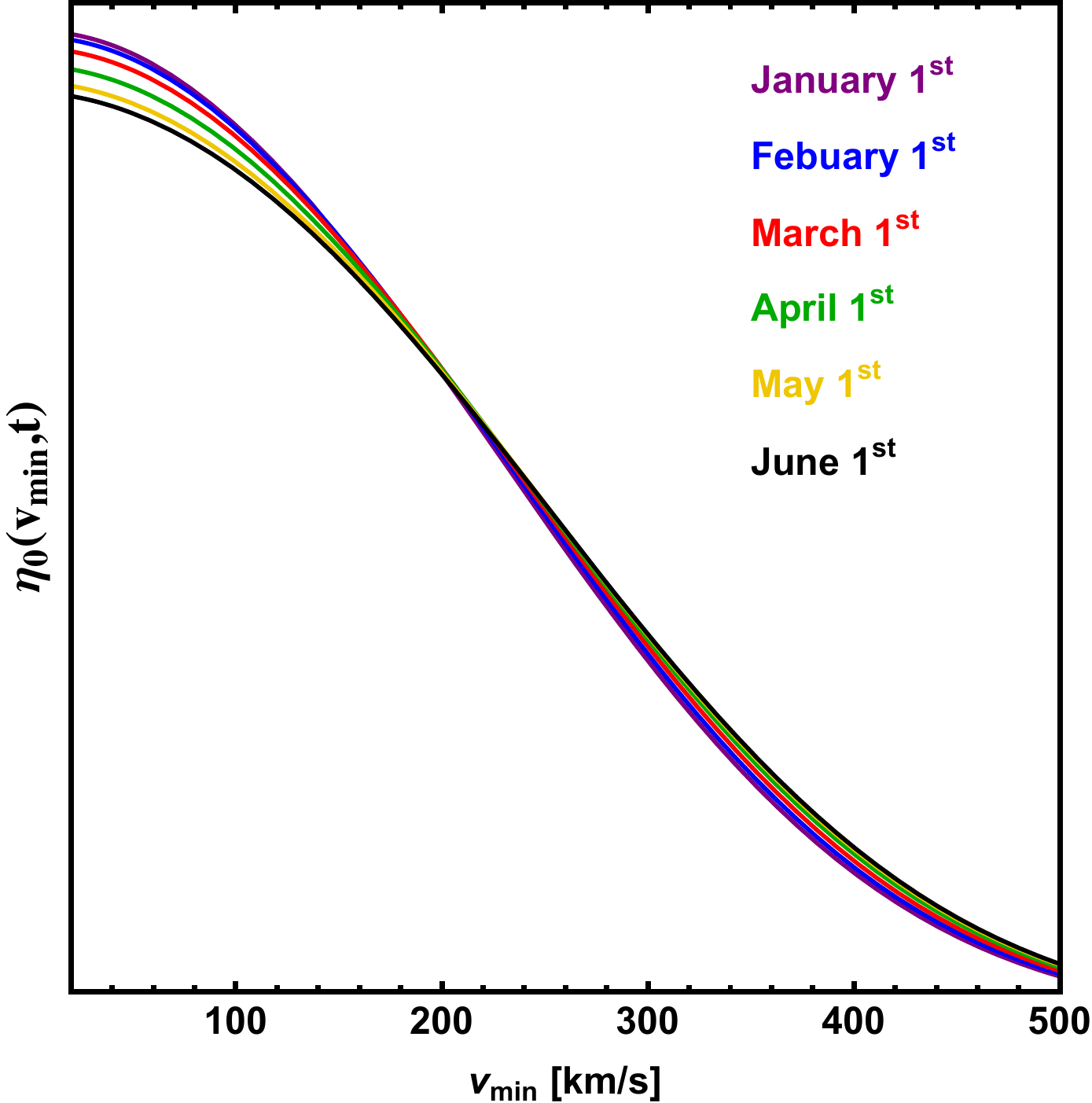}
\includegraphics[width=0.49\textwidth]{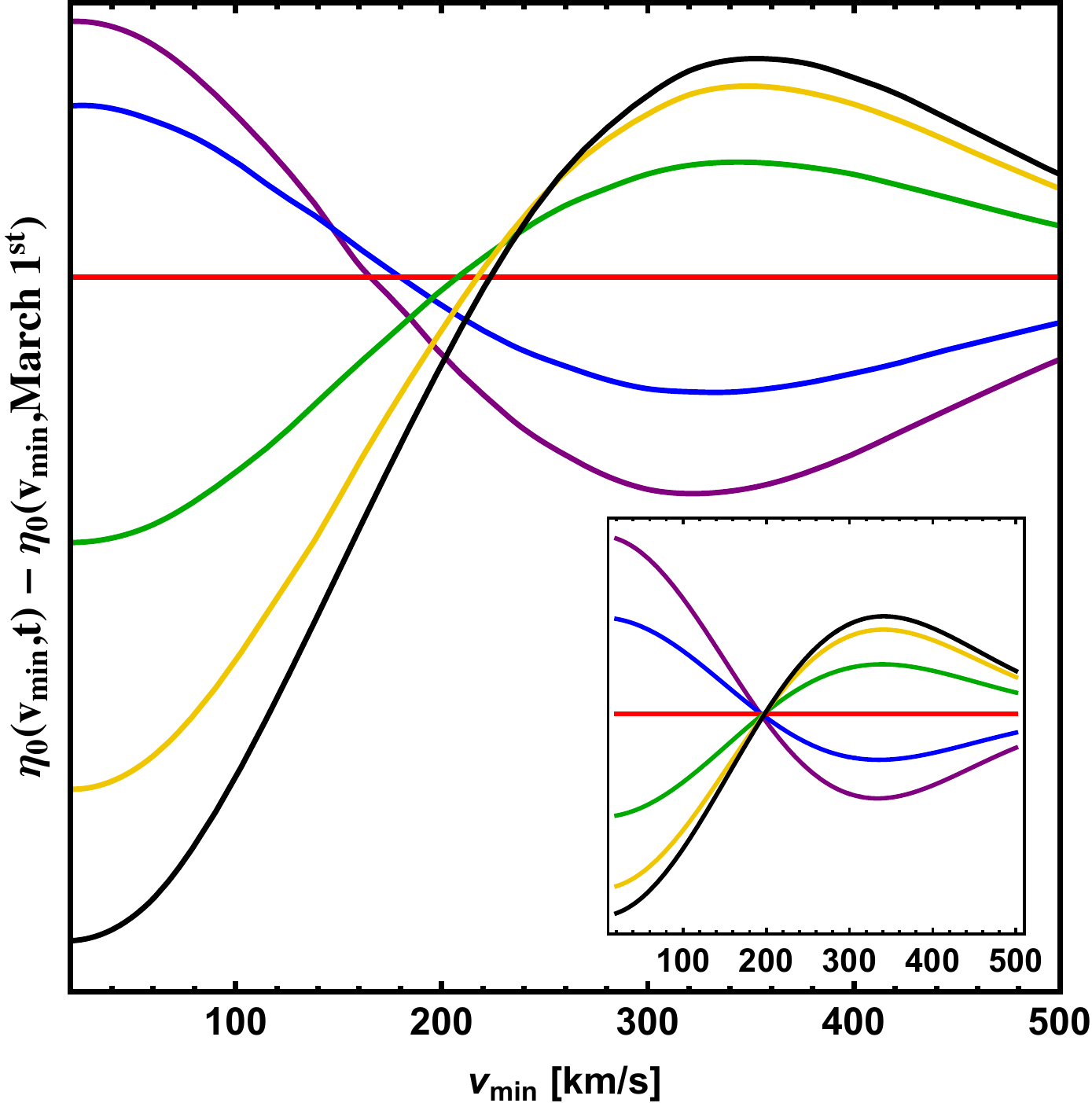}
\caption{\label{fig:eta} Left: $\eta_{0}$ plotted as a function of $\vmin$ at fixed times. Right: The difference between $\eta_{0}(\vmin,t)$ and $\eta_{0}(\vmin, t = \text{March }1^\text{st})$ evaluated at various times. The inset depicts the same figure should GF be neglected.}
\end{figure*}

Regarding point \ref{factorization}, the factorizable velocity and target dependence of the differential cross section, despite being very common, is not a completely general feature. The differential scattering cross section for DM interacting through a magnetic dipole~\cite{Pospelov:2000bq, Sigurdson:2004zp, Barger:2010gv, Chang:2010en, Cho:2010br, Heo:2009vt, Gardner:2008yn, Masso:2009mu, Banks:2010eh, Fortin:2011hv, An:2010kc, Kumar:2011iy, Barger:2012pf, DelNobile:2012tx, Cline:2012is, Weiner:2012cb, Tulin:2012uq, Cline:2012bz, Heo:2012dk, DelNobile:2013cva, Barello:2014uda, Lopes:2013xua, Keung:2010tu, Gresham:2013mua, DelNobile:2014eta, Gresham:2014vja} or an anapole moment~\cite{Pospelov:2000bq, Ho:2012bg, Fitzpatrick:2010br, Keung:2010tu, Frandsen:2013cna, Gresham:2013mua, Gao:2013vfa, DelNobile:2014eta, Gresham:2014vja} actually contains two terms with unique velocity dependences and energy-dependent coefficients. These types of differential cross sections also appear with the interactions described by some of the effective operators studied \eg in~\cite{Goodman:2010ku, Goodman:2010yf, Zheng:2010js, Liang:2013dsa, Catena:2014uqa, Catena:2014epa} (see~\cite{Fitzpatrick:2012ix, Fitzpatrick:2012ib, Kumar:2013iva, DelNobile:2013sia, Barello:2014uda} for explicit formulas of scattering amplitudes). In all these examples, velocity dependences other than the $\ud \sigma_T / \ud \ER \propto 1 / v^2$ in \Eq{diffsigmastandard} are present. This happens \eg when higher order terms in the non-relativistic (small $v$) expansion of the scattering amplitude become important. To be concrete, we can take for example the scattering rate to be
\beq
\label{r0r1rate}
\frac{\ud R_T}{\ud \ER}(\ER, t) = r_0(\ER, t) + r_1(\ER, t)
\eeq
with
\begin{align}
\label{r0r1}
r_n(\ER, t) \propto \eta_n(\vmin(\ER), t),
&&
n = 0, 1,
\end{align}
where we generalized the definition of the velocity integral in \Eq{standardeta} to
\beq\label{etan}
\eta_{n}(\vmin, t) \equiv \int_{v \geqslant \vmin} v^{2n} \, \frac{f(\bol{v}, t)}{v} \, \ud^3 v \ .
\eeq
The interesting case for us is when $r_0$ and $r_1$ have similar magnitudes. The proportionality factor between $r_i$ and $\eta_i$ in \Eq{r0r1} is in general $\ER$ dependent, and this dependence must balance the suppression provided by the extra powers of $v$ in $\eta_1$ with respect to $\eta_0$ in order for $r_0$ and $r_1$ to be comparable. We will see below that the scattering rate of a DM particle interacting through an anomalous magnetic moment has exactly this form. As is clear from \Eq{r0r1rate}, the time dependence of the rate does not coincide in general with that of a single velocity integral, as it happened instead in the simple case of \Eq{diffratestandard}. It is therefore useful to denote with $\taumax$ ($\taumin$) the time of maximum (minimum) of each velocity integral, to distinguish it from the time of maximum (minimum) of the rate denoted $\tmax$ ($\tmin$).

To understand the time-dependent behavior of $\eta_{n}$ we begin by considering the behavior of $\eta_0$. The left panel of \Fig{fig:eta} shows $\eta_0$ evaluated at the first day of the month for the first six months of the year as a function of $\vmin$. Since the behavior of the curves is difficult to discern, we plot in the right panel of \Fig{fig:eta} the difference between each of the curves in the left panel and $\eta_0$ evaluated at March $1^\text{st}$. Here, $\taumax$, the time of maximum of the velocity integral, can be seen to transition from early January to early June as $\vmin$ increases from $\approx 140$ km/s to $\approx 260$ km/s (actually, $\taumax$ occurs before January $1^\text{st}$, during the month of December at low values of $\vmin$). The inset in the right panel of \Fig{fig:eta} shows how $\eta_0$ changes with time should GF be neglected. Without GF, $\taumax$ still transitions from January $1^\text{st}$ to June $1^\text{st}$, but this transition occurs very rapidly over a very narrow range of $\vmin$ values.

\begin{figure}
\centering
\includegraphics[width=0.49\textwidth]{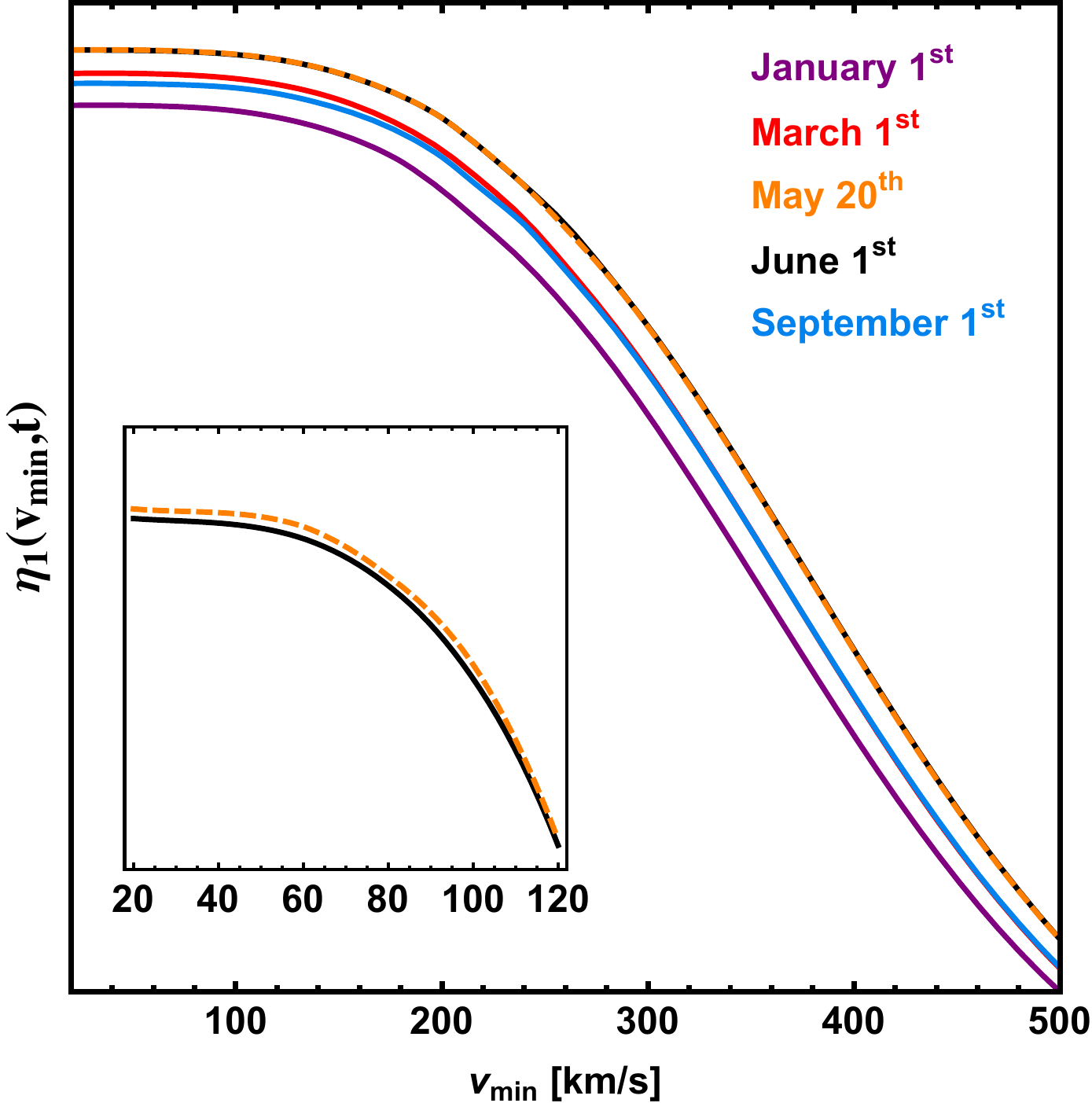}
\caption{\label{fig:eta2} $\eta_{1}$ plotted as a function of $\vmin$ at fixed times. The inset zooms in on the region where the time of maximum $\taumax$ transitions from late May, occurring at small values of $\vmin$, to early June, occurring for $\vmin \gtrsim 300$ km/s.}
\end{figure}

\Fig{fig:eta2} shows $\eta_1$ as a function of $\vmin$ for various fixed times. Unlike $\eta_0$, there appears to be a fixed separation between the various fixed time curves across nearly all values of $\vmin$. This occurs because the additional factor of $v^2$ entering the velocity integral of $\eta_1$ weights the high velocity part of the spectrum, where the fixed time curves of $\eta_0$ are visibly separated. The inset of \Fig{fig:eta2} zooms in on the low $\vmin$ region to emphasize that $\taumax$ of $\eta_1$ does have a small $\vmin$ dependence, transitioning from late May at small values of $\vmin$, to early June for $\vmin \gtrsim 300$.

\begin{figure*}[t!]
\centering
\includegraphics[width=\textwidth, trim=0cm 1.1cm 0cm 1.2cm, clip=true]{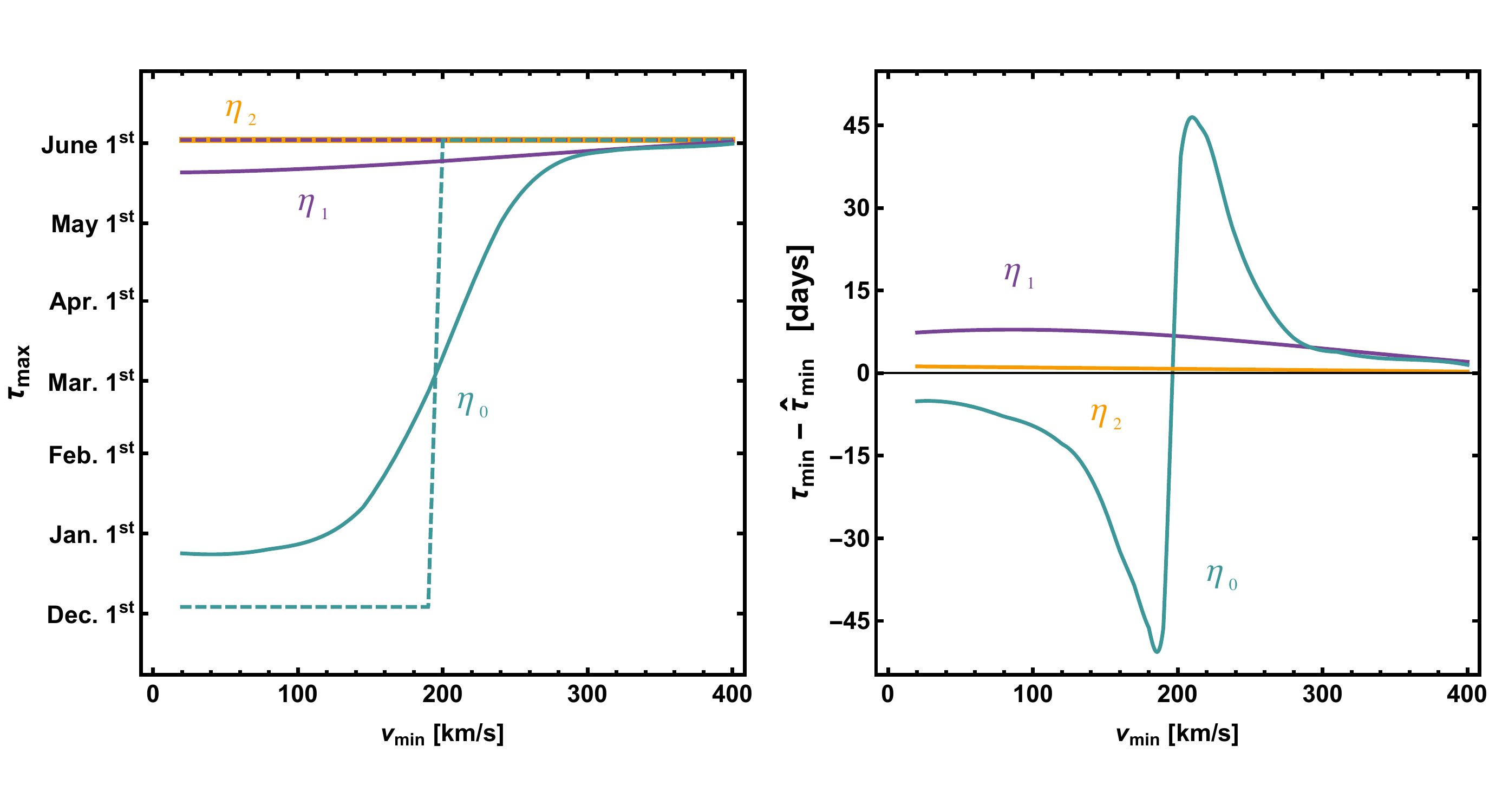}
\caption{\label{fig:SHM} Time of maximum $\taumax$ (left) and minimum $\taumin$ (right) of $\eta_{0}$, $\eta_{1}$, and $\eta_{2}$ assuming the SHM, with (solid) and without (dashed) GF. In the right panel we plot $\taumin - \htaumin$, where $\htaumin$ is $\taumax + 6$ months. Neglecting GF, $\taumin$ is nearly indistinguishable from $\htaumin$, and thus is not shown.}
\end{figure*}

For $n > 1$, one would expect the high end of the velocity distribution to become increasingly weighted, which within the SHM should result in a time dependence similar to that of $\eta_1$, but even more independent of $\vmin$. This is shown in \Fig{fig:SHM}, where $\taumax$ and $\taumin$ are plotted for $\eta_0$, $\eta_1$, and $\eta_2$. Instead of plotting $\taumin$, we plot $\taumin - \htaumin$, where $\htaumin \equiv \taumax + 6$ months. \Fig{fig:SHM} shows the effect of including (solid) and neglecting (dashed) GF.\footnote{Unless otherwise stated, GF and the eccentricity of Earth's orbit are included in all calculations.} For $\eta_2$, $\taumax$ is hardly affected by GF and thus only a single solid line is plotted. The results for $\taumin - \htaumin$ without GF are not shown as in this case $\taumin$ is nearly indistinguishable from $\htaumin$. 

\Fig{fig:SHM} shows that, within the SHM, $\eta_0$ is the only $\eta_n$ whose time-dependent behavior differs markedly from $\eta_{n\geqslant 1}$. Thus, for the target-dependent features of the modulation to appear, assuming no other forms of anisotropy are present within the dark halo, the differential cross section must not only contain a non-factorizable velocity dependence, but one of the terms in the differential cross section must be proportional to $\eta_0$. Should $\taumax$ and $\taumin$ of $\eta_0$ become $\vmin$ dependent above $300$ km/s, \eg due to the presence of DM substructure~\cite{DelNobile:2015nua}, the approximate degeneracy of $\eta_{n\geqslant 1}$ (and near exact degeneracy of $\eta_{n \geqslant 2}$) would break and the previous requirement would no longer be necessary.

We would like to note that any $\eta_n$ can actually be rewritten in terms of, and thus computed from, $\eta_0$. Defining $F(v, t) \equiv v^2 \int \ud\Omega \, f(\bol{v}, t)$ with $\ud^3 v = v^2 \, \ud v \, \ud \Omega$, one can write
\beq
\eta_n = \int_{v \geqslant \vmin} v^{2n} \, \frac{f(\bol{v}, t)}{v} \, \ud^3 v = \int_{\vmin}^{\infty} v^{2n} \frac{F(v, t)}{v} \, \ud v \ ,
\label{etapart1}
\eeq
which implies
\beq
\eta_n = - \int_{\vmin}^{\infty} v^{2n} \frac{\ud \eta_{{0}}(v, t)}{\ud v} \, \ud v \ ,
\label{etapart2}
\eeq
as can be seen by differentiating \Eq{standardeta}. Finally, integrating \Eq{etapart2} by parts yields
\beq
\eta_{n}(\vmin,t) = \vmin^{2n} \eta_{0}(\vmin, t) + 2 n \int_{\vmin}^{\infty} v^{2n-1} \eta_{0}(v, t) \, \ud v \ ,
\label{etapartthree}
\eeq
where we used the fact that $\eta_0(\infty, t) = 0$. With a similar set of manipulations, any arbitrary $\eta_n$ can be written in terms of any other arbitrary $\eta_{n'}$. Therefore, in principle, one may choose to express the rate in terms of any of the $\eta_n$ (or even in terms of $f(\bol{v}, t)$ itself, as shown in Eq.~(18) of~\cite{DelNobile:2013cva}). Some of the $\eta_n$ may have good properties for specific calculations, for example the normalization condition $\int f(\bol{v}, t) \, \ud^3 v = 1$ can be written either as $\int_0^\infty \eta_0(\vmin, t) \, \ud \vmin = 1$ (see \eg\cite{Feldstein:2014ufa,DelNobile:2015uua}) or $\eta_{\frac{1}{2}}(0) = 1$. Moreover, whenever the velocity integrals need to be computed numerically (\eg for complicated halo models, or when computing the effect of GF), \Eq{etapartthree} can be used to straightforwardly determine $\eta_{n \neq 0}$ once $\eta_0$ is known.

The different time dependence of the the various $\eta_n$ can be understood by looking at \Eq{etapartthree}. Were it only for the first term on the right-hand side, $\eta_{n \neq 0}$ and $\eta_0$ would obviously have the same time dependence at fixed $\vmin$. Because of the second term, however, $\eta_n(\vmin, t)$ is a function of time that depends in a nontrivial way on $\eta_0(v, t)$ for all $v \geqslant \vmin$.

\section{Annual modulation for magnetic dipole DM}\label{MDMover}
\subsection{Elastic scattering}
\label{mdm}

\begin{figure*}
\centering
\includegraphics[width=\textwidth]{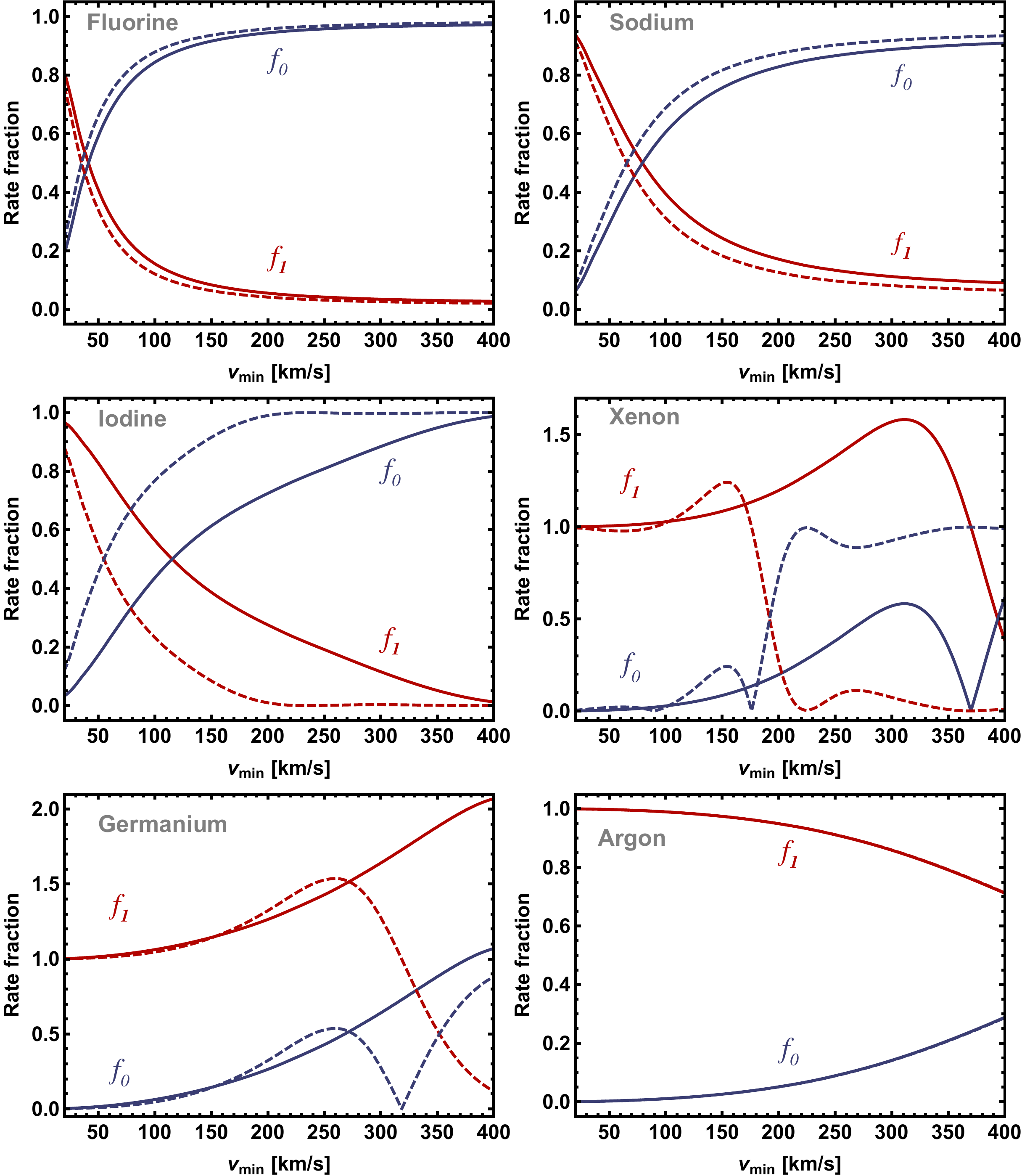}
\caption{\label{fig:ratefrac} Rate fractions $f_0$ and $f_1$, as defined in \Eq{f0f1}, for fluorine (top left), sodium (top right), iodine (middle left), xenon (middle right), germanium (bottom left), and argon (bottom right). Solid (dashed) lines correspond to $m = 100$ GeV ($1$ TeV).}
\end{figure*}

We study here in detail the case of a Dirac fermion DM candidate $\chi$ elastically scattering with nuclei through an anomalous magnetic dipole moment $\lambda_\chi$, with interaction Lagrangian $\Lag = (\lambda_\chi/2) \, \bar{\chi} \sigma_{\mu \nu} \chi F^{\mu \nu}$. The differential cross section for elastic scattering off a target nuclide $T$ with $Z_T$ protons and spin $S_T$ is
\begin{multline}
\label{diffsigmamagnetic}
\frac{\ud \sigma_T}{\ud \ER}(\vmin, v) =
\alpha \lambda_\chi^2 \left\{ Z_T^2 \frac{m_T}{2 \mu_T^2} \left[ \frac{1}{\vmin^2} - \frac{1}{v^2} \left( 1 - \frac{\mu_T^2}{m^2} \right) \right] F_{\text{SI}, T}^2(\ER(\vmin)) + \right. \\  \left. \frac{\hat\lambda_T^2}{v^2} \frac{m_T}{m_p^2} \left( \frac{S_T + 1}{3 S_T} \right) F_{\text{M}, T}^2(\ER(\vmin)) \right\} ,
\end{multline}
with $\alpha = e^2/4\pi$ the electromagnetic fine structure constant, $m_p$ the proton mass, $\hat{\lambda}_T$ the nuclear magnetic moment in units of the nuclear magneton $e/(2m_p) = 0.16$ $\text{GeV}^{-1}$, and $\ER(\vmin) = 2 \mu_T^2 \vmin^2 / m_T$~\cite{DelNobile:2013cva}. The differential cross section contains two terms, one arising from the charge-dipole interaction and the other arising from the dipole-dipole interaction. The former thus depends on the nuclear charge and contains a spin-independent form factor while the latter depends on the nuclear spin and contains a magnetic form factor. Both form factors are normalized to $1$ at zero recoil energy. We compute the cross section with the formalism and form factors provided in~\cite{Fitzpatrick:2012ix, Fitzpatrick:2012ib}.

\begin{figure*}[t!]
\centering
\includegraphics[width=0.49\textwidth]{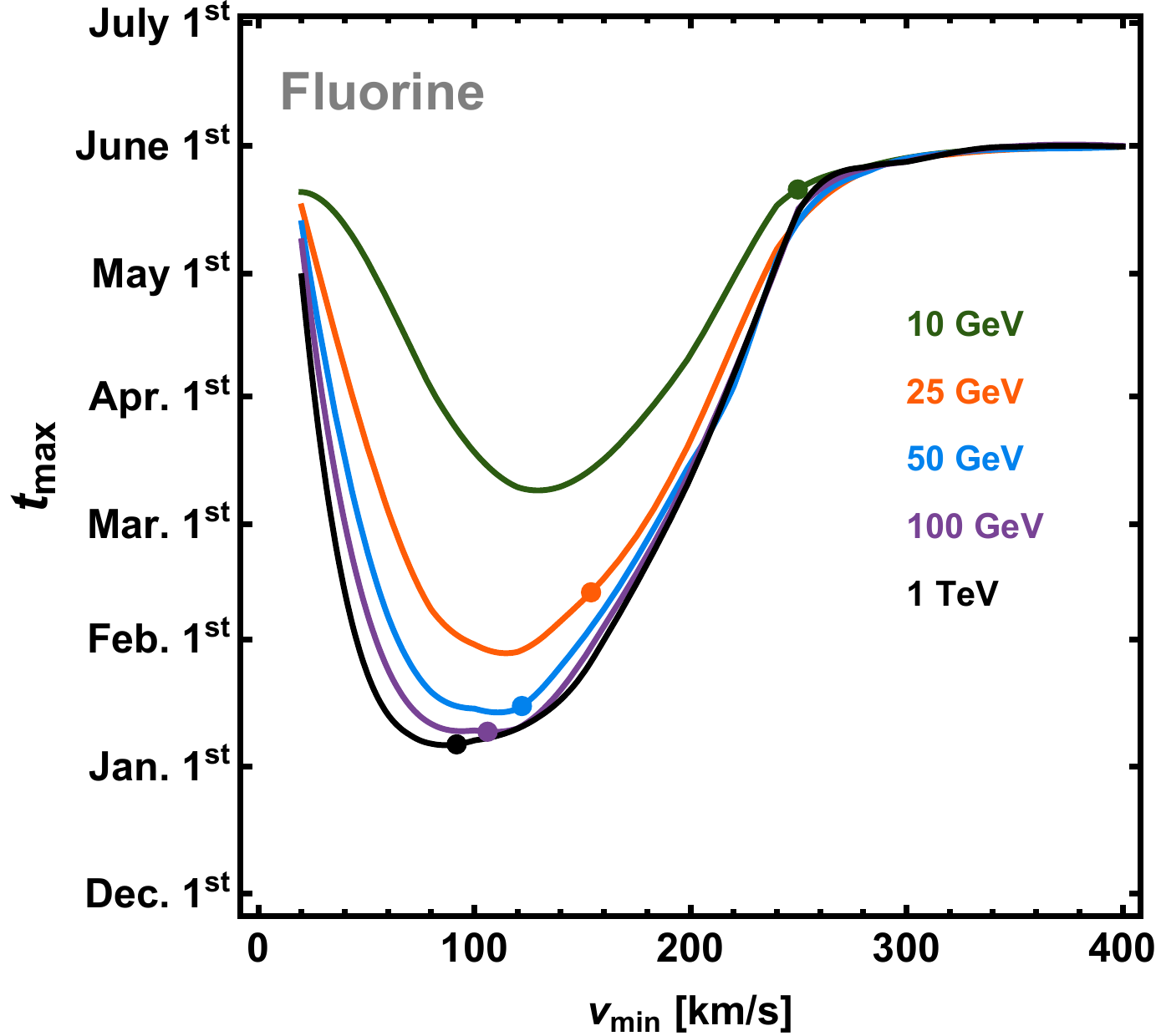}
\includegraphics[width=0.49\textwidth]{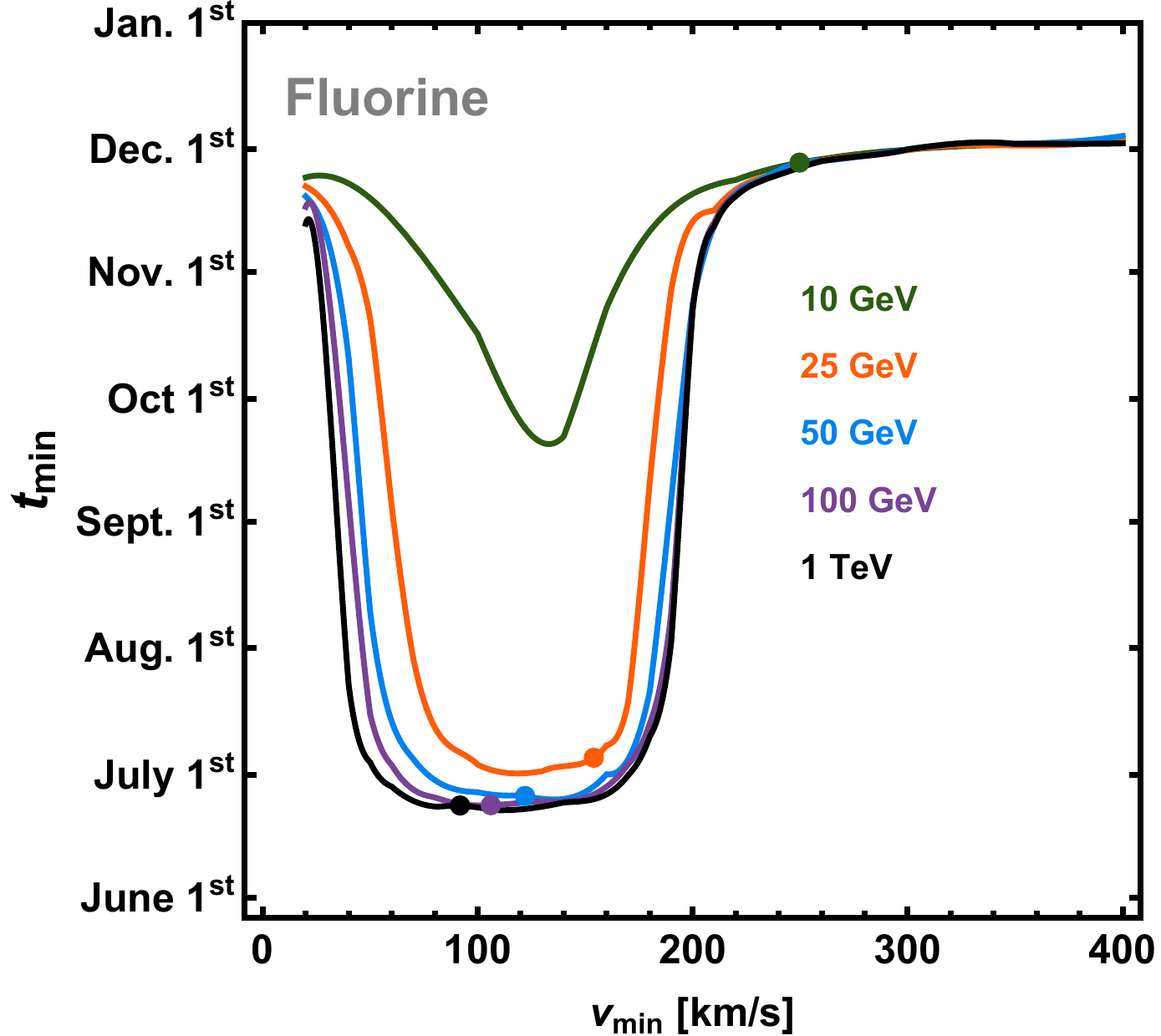}
\caption{\label{fig:Fmaxandmin} Time of maximum $\tmax$ (left) and minimum $\tmin$ (right) of the differential rate for magnetic DM scattering off fluorine, plotted for various DM masses as a function of $\vmin$. The current low energy threshold for PICO has been mapped onto $\vmin$ for each DM mass and is shown as a small solid dot.}
\end{figure*}

Since the magnetic DM differential cross section contains terms proportional to $\eta_{0}(\vmin, t)$ and $\eta_{1}(\vmin, t)$, the modulation of the differential rate is a direct consequence of the interplay of these two functions and their respective coefficients. The relative importance of each of these functions is determined by the target and DM mass-dependent coefficients. We define $r_0(\ER, t)$ and $r_1(\ER, t)$ as the terms of the differential rate containing $\eta_{0}$ and $\eta_{1}$ respectively, as in Eqs.~\eqref{r0r1rate} and \eqref{r0r1}, and $\bar{r}_0(\ER)$ and $\bar{r}_1(\ER)$ to be their time average. The time-averaged differential rate reads then $\ud \bar{R}_T(\ER) / \ud \ER = \bar{r}_0 + \bar{r}_1$. Fig.~\ref{fig:ratefrac} depicts the absolute value of the time-averaged rate fractions,
\begin{align}
\label{f0f1}
f_0 \equiv \frac{|\bar{r}_0|}{\bar{r}_0 + \bar{r}_1} \ ,
&&&
f_1 \equiv \frac{|\bar{r}_1|}{\bar{r}_0 + \bar{r}_1} \ ,
\end{align}
as functions of $\vmin$, for six elements (fluorine, iodine, sodium, xenon, germanium, and argon) employed by current DM direct detection experiments. When more than one isotope is present, \ie for germanium and xenon, $r_0$ and $r_1$ are understood to be summed over isotopes. Solid (dashed) lines correspond to a $100$ GeV ($1$ TeV) DM particle.

\begin{figure*}[t!]
\centering
\includegraphics[width=0.49\textwidth]{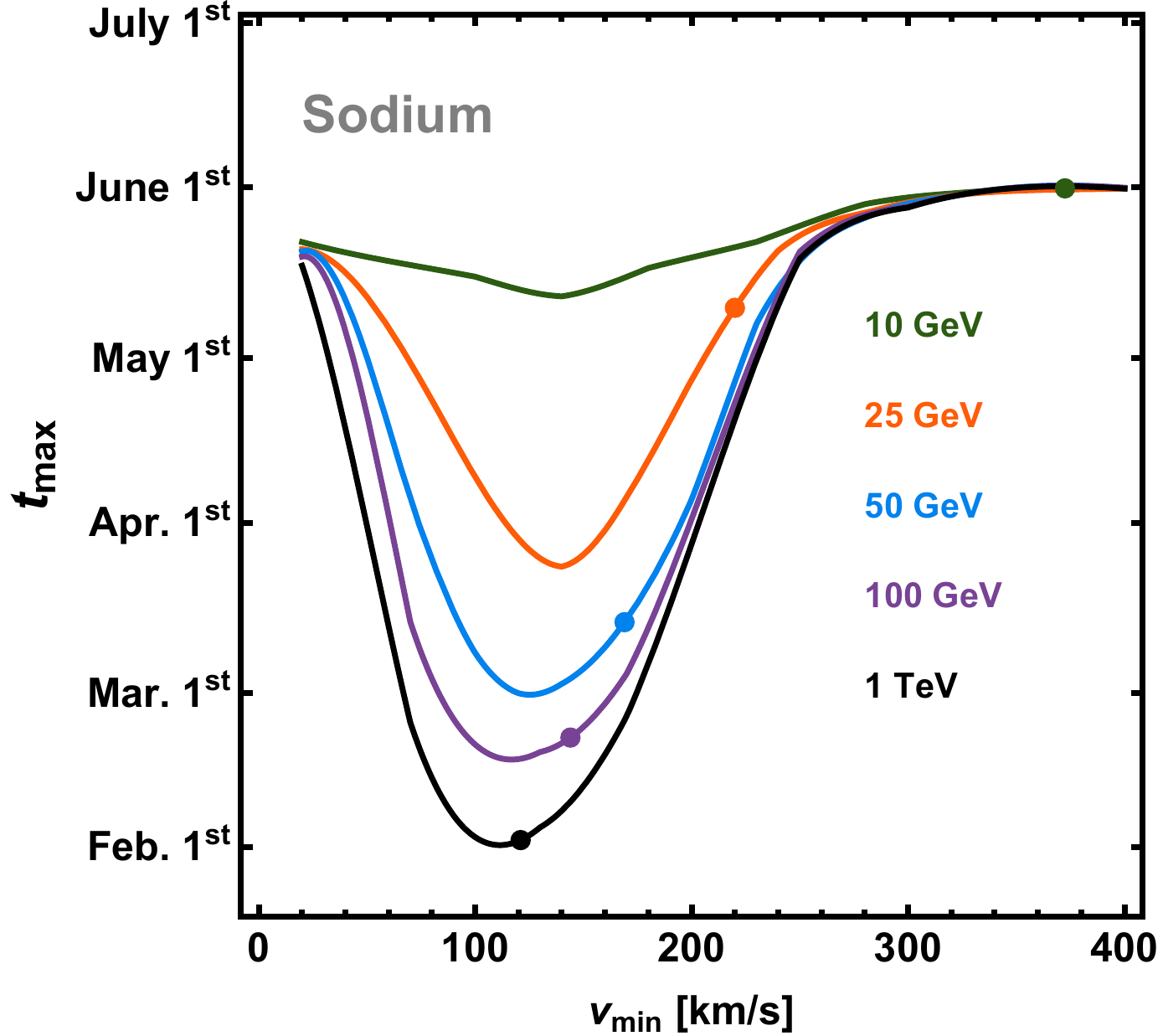}
\includegraphics[width=0.49\textwidth]{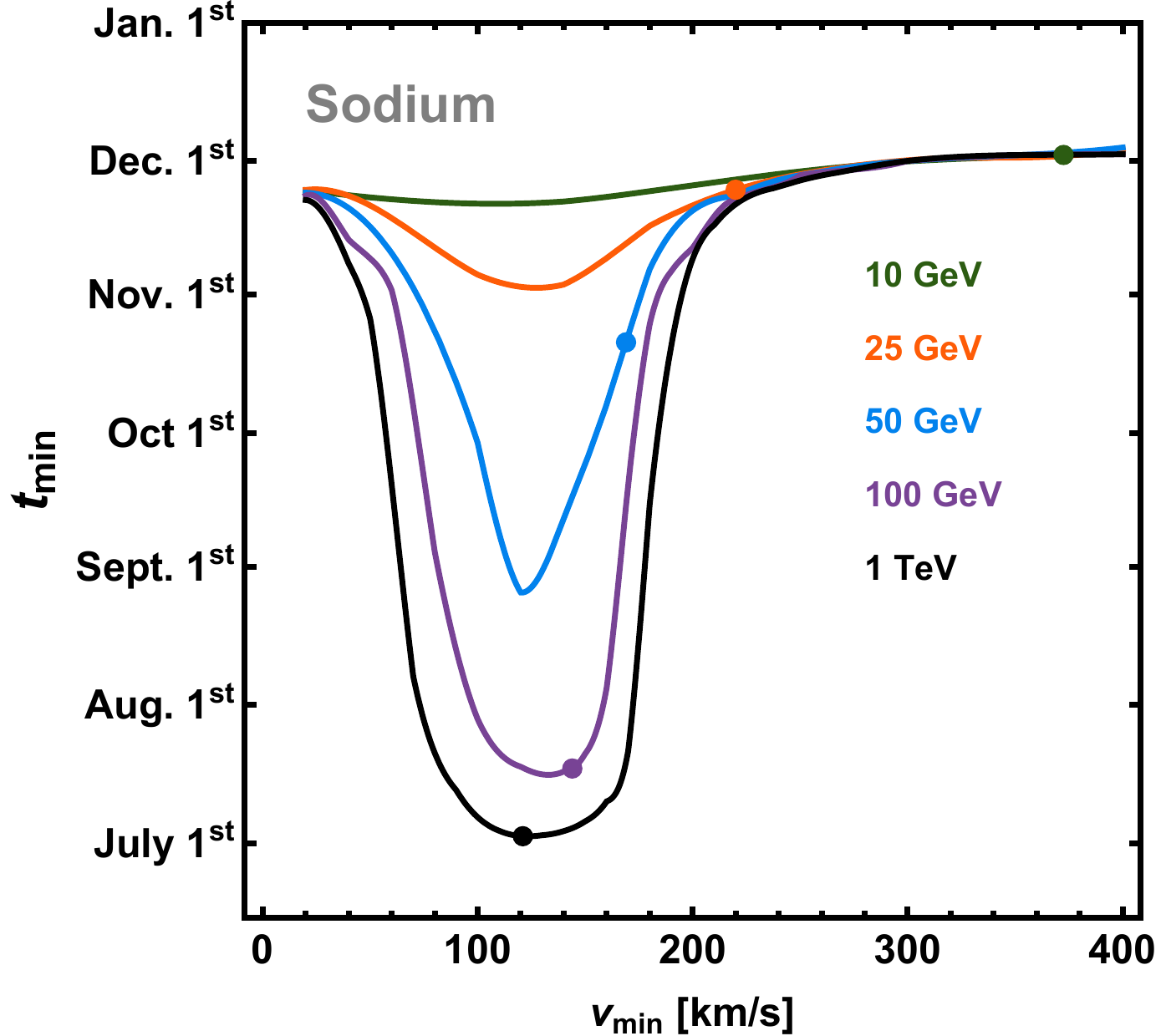}
\caption{\label{fig:Namaxandmin} Time of maximum $\tmax$ (left) and minimum $\tmin$ (right) of the differential rate for magnetic DM scattering off sodium, plotted for various DM masses as functions of $\vmin$. The current low energy threshold for DAMA has been mapped onto $\vmin$, assuming a quenching factor $Q_\text{Na}=0.3$, for each DM mass and is shown as a small solid dot.}
\end{figure*}

\begin{figure*}[t!]
\centering
\includegraphics[width=0.49\textwidth]{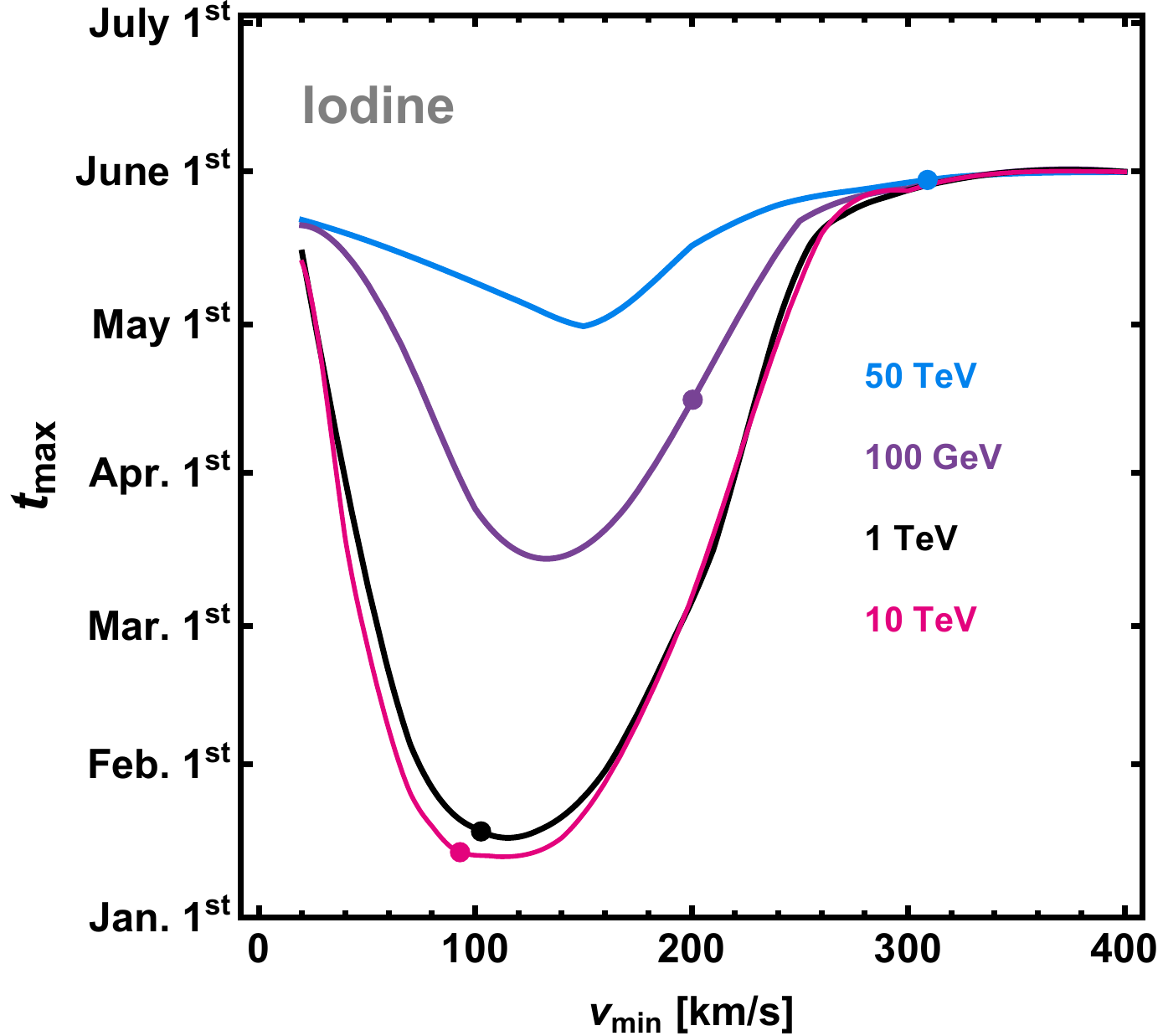}
\includegraphics[width=0.49\textwidth]{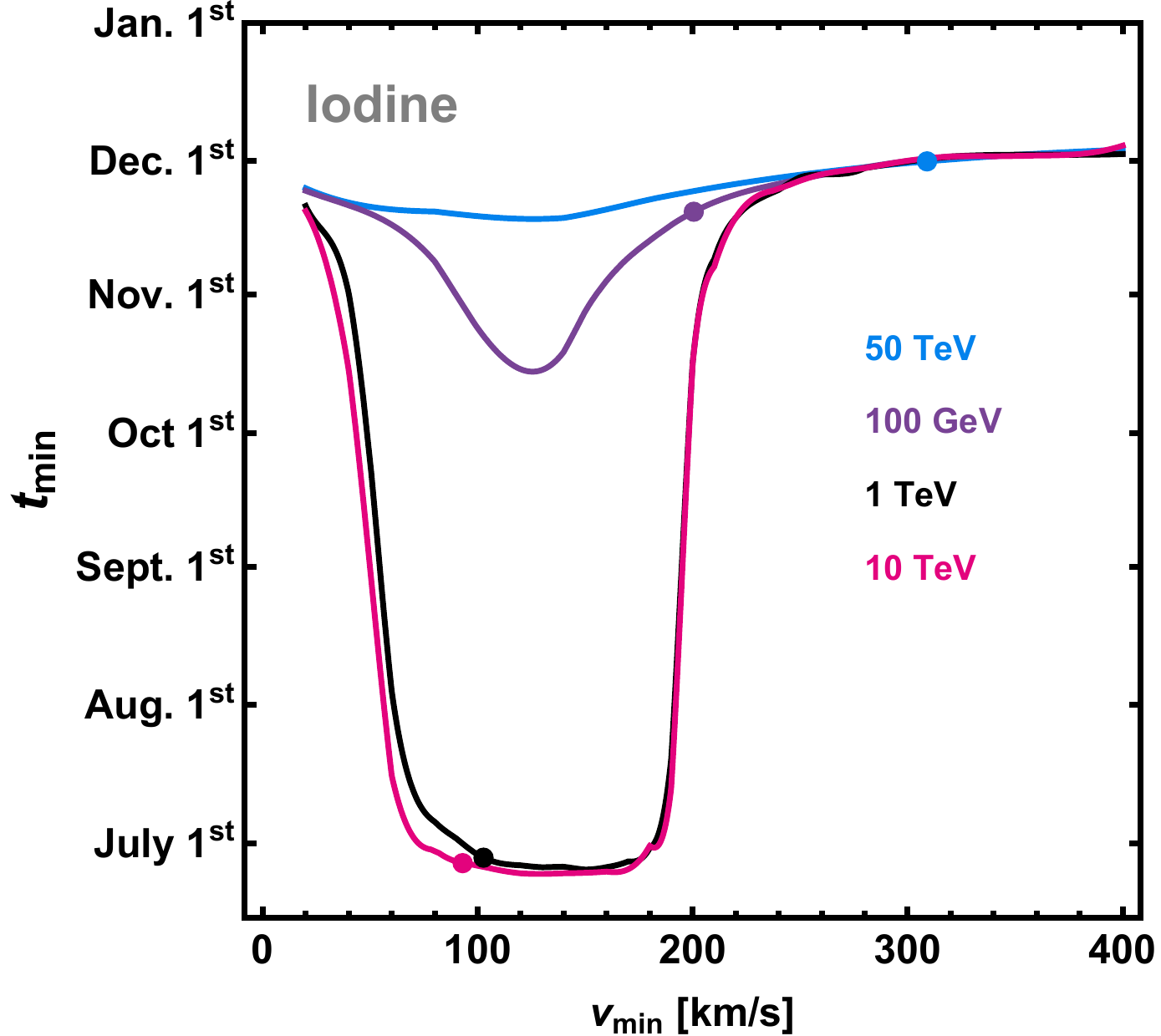}
\caption{\label{fig:Imaxandmin} Time of maximum $\tmax$ (left) and minimum $\tmin$ (right) of the differential rate for magnetic DM scattering off iodine, plotted for various DM masses as functions of $\vmin$. The current low energy threshold for DAMA has been mapped onto $\vmin$, assuming a quenching factor $Q_\text{I}=0.09$, for each DM mass and is shown as a small solid dot.}
\end{figure*}

\begin{figure*}[t!]
\centering
\includegraphics[width=0.49\textwidth]{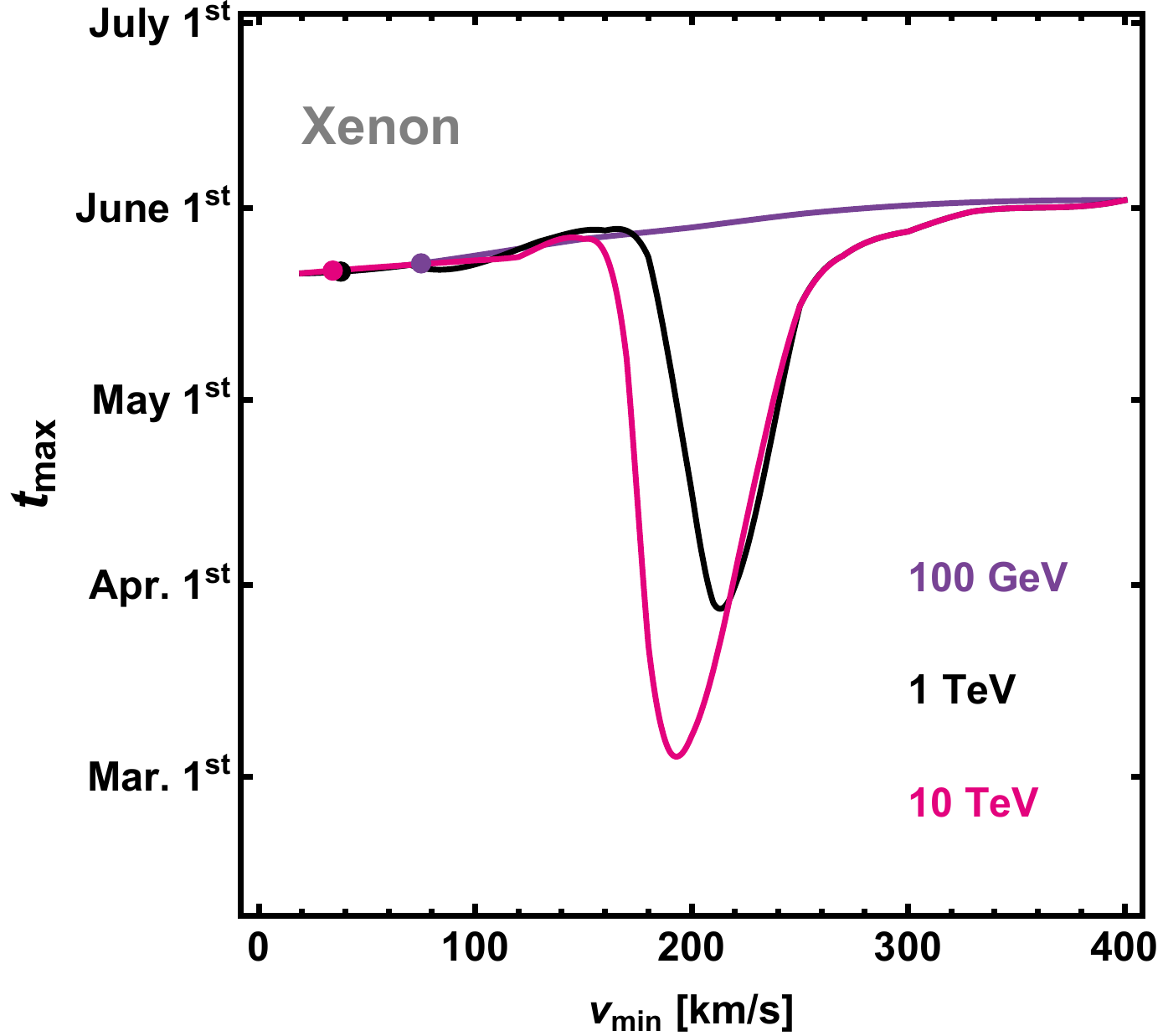}
\includegraphics[width=0.49\textwidth]{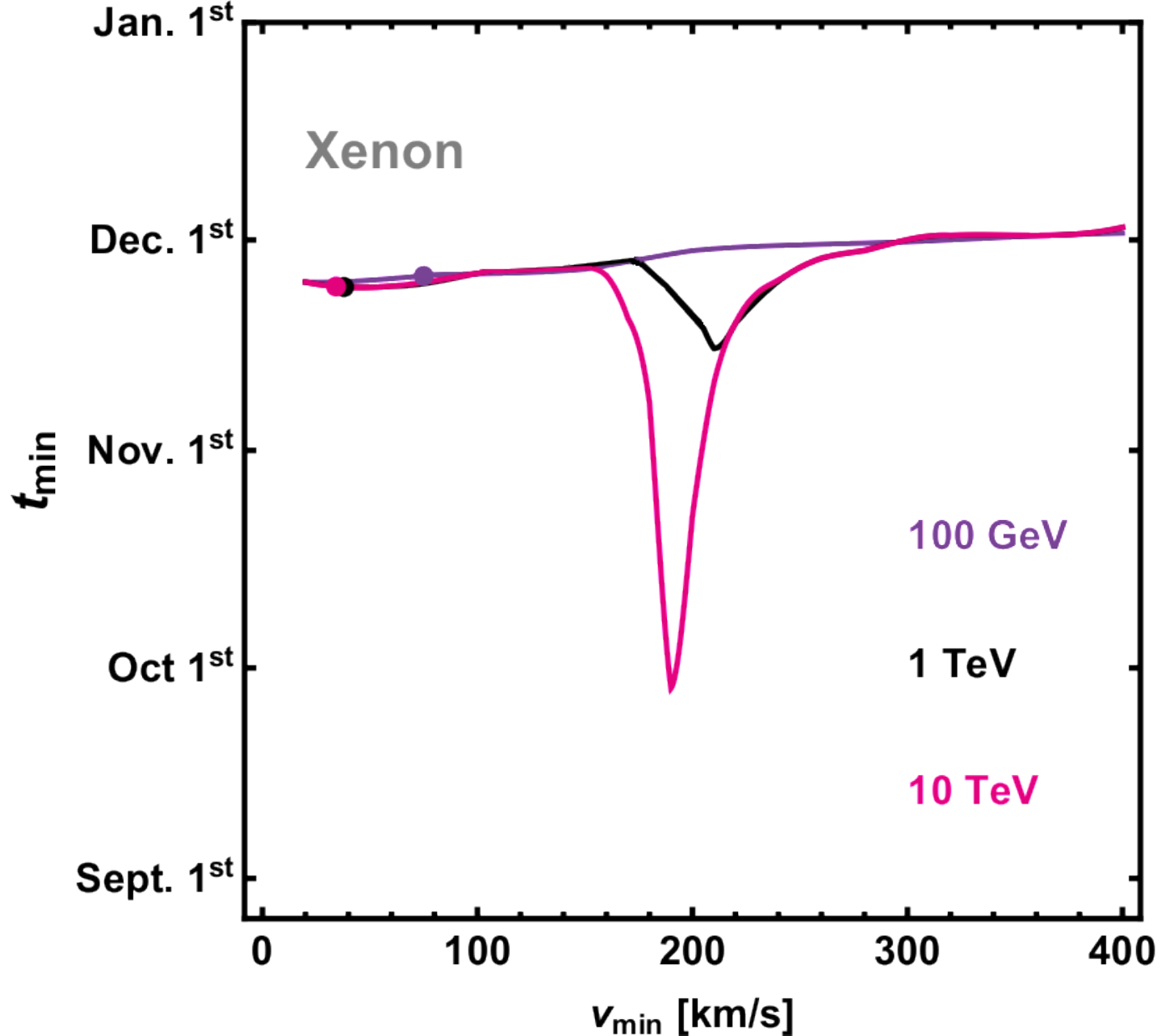}
\caption{\label{fig:Xemaxandmin} Time of maximum $\tmax$ (left) and minimum $\tmin$ (right) of the differential rate for magnetic DM scattering off xenon, plotted for various DM masses as functions of $\vmin$. The current low energy threshold for LUX has been mapped onto $\vmin$ for each DM mass and is shown as a small solid dot. }
\end{figure*}

The target dependence of $\tmax$ and $\tmin$ can be understood by combining the information on the time dependence of $\eta_{0}$ and $\eta_{1}$ in \Fig{fig:SHM} with the information on the rate fraction of the corresponding element shown in Fig.~\ref{fig:ratefrac}. $\tmax$ and $\tmin$ as functions of $\vmin$ are shown in Figs.~\ref{fig:Fmaxandmin}--\ref{fig:Xemaxandmin} for magnetic DM scattering off fluorine, sodium, iodine, and xenon. We have chosen not to plot $\tmax$ and $\tmin$ for germanium and argon because the results for all DM masses below $10$ TeV are identical due to their small (germanium) or zero (argon) nuclear magnetic moment (see Ref.~\cite{DelNobile:2015tza} for details). For each element, $\tmax$ (left panels) and $\tmin$ (right panels) are plotted for various DM masses ranging from $10$ GeV to $10$ TeV. Also shown, depicted as dots on the $\tmax$ and $\tmin$ curves, are the $\ER$ thresholds for LUX~\cite{Akerib:2013tjd} ($3.1$ keV~\cite{DelNobile:2015uua}, employing Xe), DAMA~\cite{Bernabei:2013xsa} ($6.7$ keV for Na with $Q_\text{Na} = 0.3$ and $22.2$ keV for I with $Q_\text{I} = 0.09$), and PICO~\cite{Amole:2015lsj} ($3.2$ keV, employing F), translated into $\vmin$ for elastic scattering with each DM mass. When multiple isotopes are present, the value of $m_T$ in \Eq{vmin} is replaced with $\sum_T \xi_T m_T$, where $\xi_T$ is the numerical abundance of element $T$.

Figs.~\ref{fig:Fmaxandmin}--\ref{fig:Xemaxandmin} show that $\tmax$ and $\tmin$ become target and DM mass independent for $\vmin \gtrsim 300$ km/s. This is due to the fact that the difference in the time-dependent behavior between $\eta_{0}$ and $\eta_{1}$, which are central to the target-dependent features, vanish above $\vmin \approx 300$ km/s (see \Fig{fig:SHM}), if the only source of anisotropy in the local halo is GF. 

Fig.~\ref{fig:ratefrac} confirms that at sufficiently small values of $\vmin$ the contribution to the differential rate from the term proportional to $\eta_{0}$ can be neglected. This is because the $r_{1}$ term contains the factor $1 / \vmin^2$, which dominates the $\vmin$ dependence of the rate at small $\vmin$ values. Thus, in the small $\vmin$ limit, $\tmax$ occurs in late May and $\tmin$ occurs in late November, regardless of the target element and DM mass. This behavior is a feature of elastic magnetic DM and other DM models could behave in a qualitatively different way.

For target elements that have a nonzero average nuclear magnetic moment (i.e., all elements considered here except argon), at large enough values of $\vmin$ the dipole-dipole interaction inevitably becomes dominant, and thus $r_{0} > r_{1}$. This is because the spin-independent form factor in \Eq{diffsigmamagnetic} decreases significantly faster than the magnetic form factor. Fig.~\ref{fig:ratefrac} confirms that for all elements considered except argon, there exists a value of $\vmin$ below which $r_{1}$ is the dominant contribution to the rate, and above which $r_{0}$ is the dominant contribution to the rate. The location in $\vmin$ of this transition and how fast or gradual it is determine the unique element-dependent features of $\tmax$ and $\tmin$ in Figs.~\ref{fig:Fmaxandmin}--\ref{fig:Xemaxandmin}.

The mass of the DM particle can have a large influence on the appearance of target-dependent features. Consider for instance the difference between a $100$ GeV and $1$ TeV DM particle scattering off xenon. For a $100$ GeV DM particle, \Fig{fig:ratefrac} shows that the $\vmin$ point at which $r_{0}$ becomes dominant is around $\vmin \approx 400$ km/s. Since this value of $\vmin$ lies in the target-independent region, $\tmax$ is effectively determined solely by the time dependence of $\eta_{1}$. As the DM mass increases, the point at which $r_{0}$ becomes dominant with respect to $r_{1}$ shifts to lower values of $\vmin$. This is partly due to the fact that the $\vmin$ value corresponding to a given $\ER$ decreases, but also because the terms $1/\mu_T^2$ and $\mu_T^2/m^2$ multiplying the SI component of \Eq{diffsigmamagnetic} decrease. Consequently, for a $1$ TeV DM particle scattering off xenon, the $\vmin$ value at which $r_{0}$ becomes dominant appears in a $\vmin$ region where the time dependence of $\eta_{1}$ and $\eta_{0}$ differ, leading to the appearance of a unique target-dependent feature in the $\tmax$ and $\tmin$ curves.

Up to this point we have only discussed how target-dependent modulation arises and how, under the assumption of magnetic DM, observables associated with the modulation of the rate in $\vmin$ can potentially change. We have not yet discussed how these effects would manifest in present day experiments. To determine if experiments are capable of observing these target-dependent features, one must take into account the experimental threshold, the efficiency, the energy resolution, and the binning method.

The obvious requirement for these target-dependent effects to be observable, is that the experimental threshold in $\vmin$ must be below $300$ km/s. The threshold in $\vmin$ depends on the threshold in $\Ed$, the DM particle mass, and the scattering kinematics. Figs.~\ref{fig:Fmaxandmin} and \ref{fig:Xemaxandmin} show that present thresholds are already low enough to give rise to a four month difference in $\tmax$ for a $50$ GeV DM particle scattering elastically off fluorine and xenon (while the $50$ GeV curve is not shown for xenon, it directly overlaps with the $100$ GeV curve), should the differential scattering rate be measured with perfect energy resolution, which is not possible for actual experiments.

\begin{figure}[t!]
\centering
\includegraphics[width=0.49\textwidth]{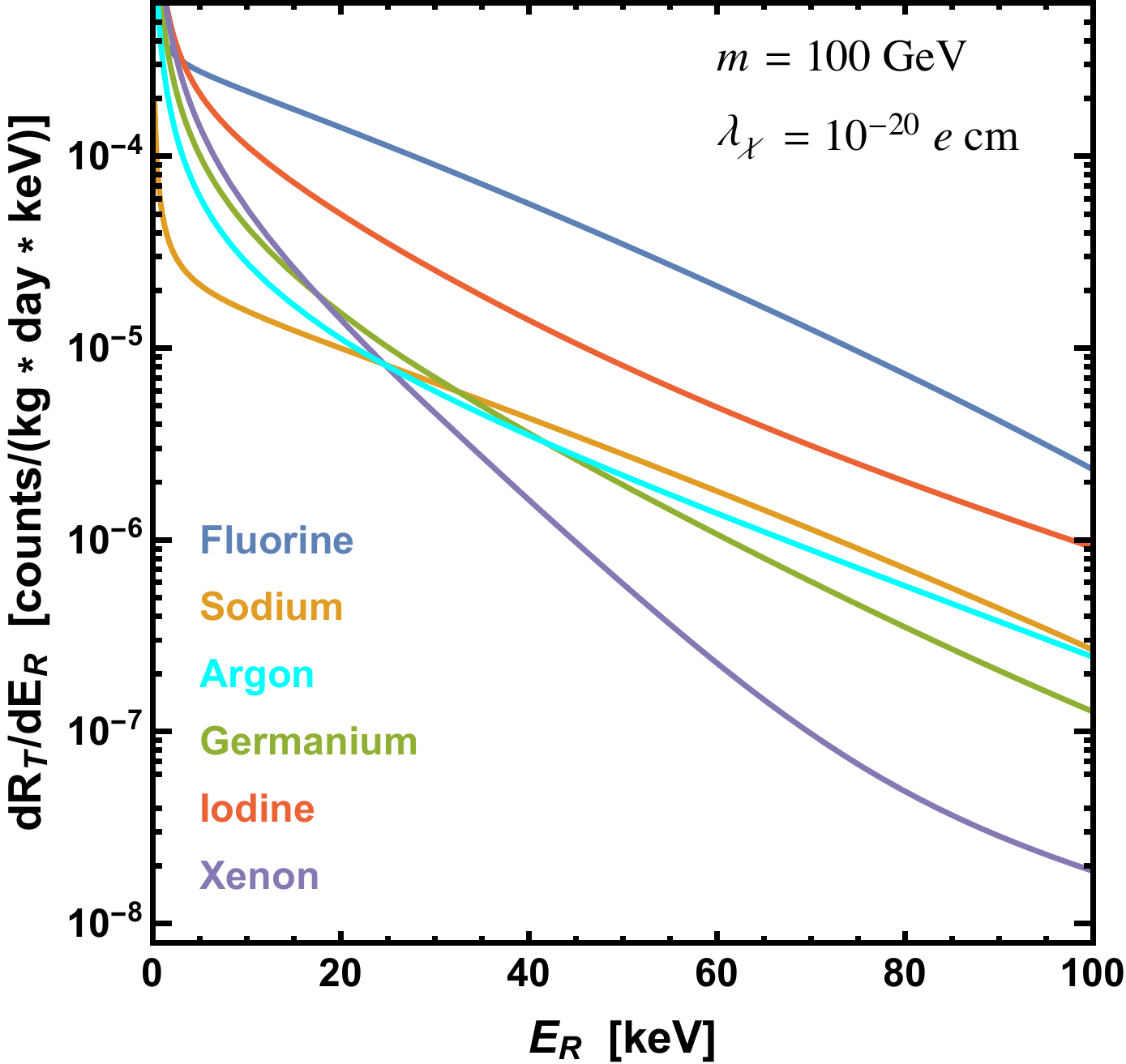}
\includegraphics[width=0.49\textwidth]{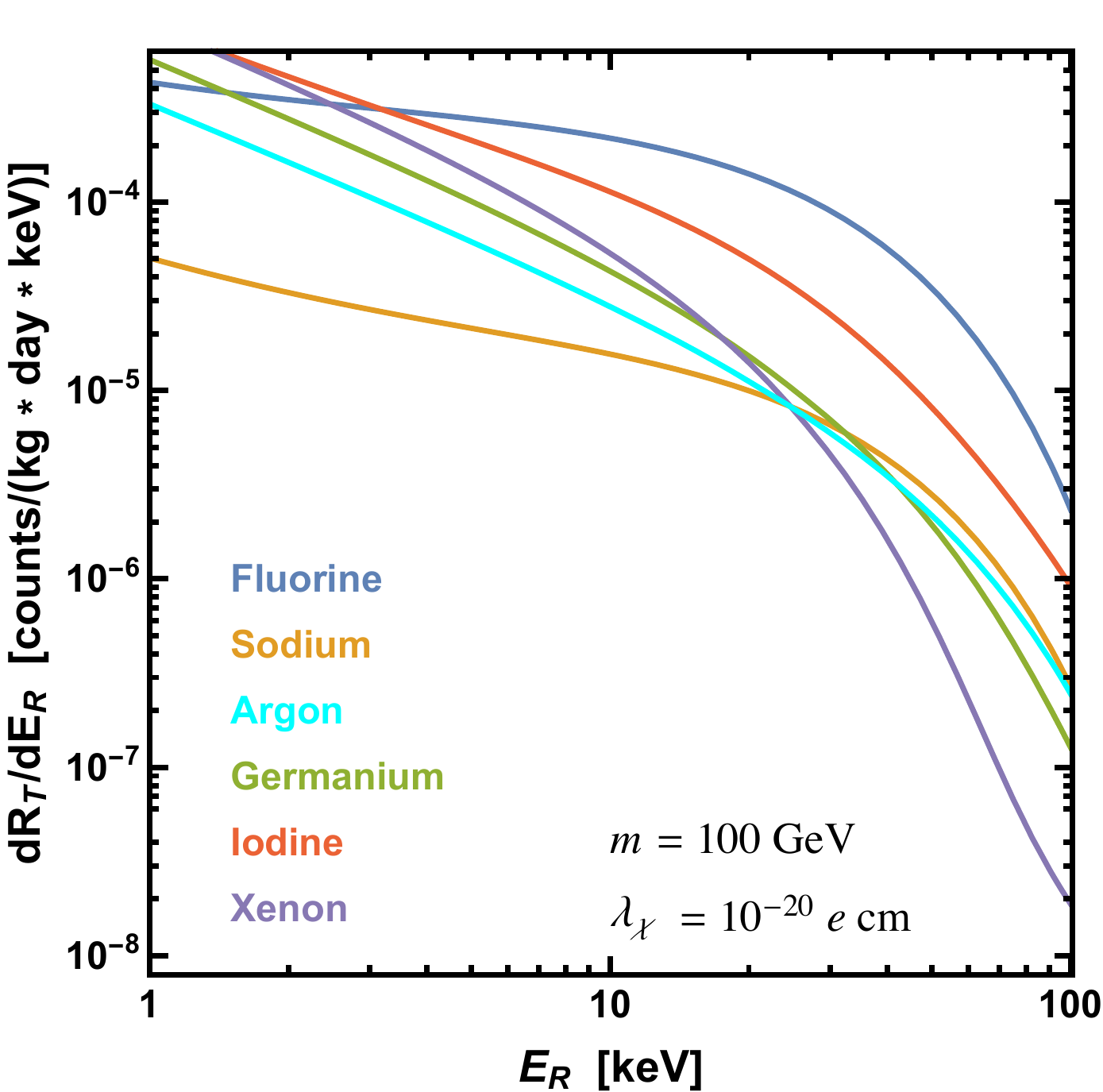}
\caption{\label{fig:diffrate} The time-averaged differential rate (summed over isotopes) in units of counts/(kg day keV) for a $100$ GeV magnetic DM particle scattering off various elements as a function of recoil energy, shown on a semi-log (left) and log-log (right) plot. $\lambda_\chi$ has been set to $10^{-20} \, e $ cm. }
\end{figure}

Since we would like to see how observable this target dependence could be, we choose to consider experiments employing elements with large nuclear magnetic moments. For this reason we begin by considering the fluorine-based experiment PICO. PICO measures the energy-integrated rate as a function of threshold energy $E_\text{th}$, and has an energy-dependent efficiency function that reduces the contribution of the scattering events near threshold. Figs.~\ref{fig:Fmaxandmin} and \ref{fig:diffrate} can be used to understand how much the modulation features in the differential rate are erased in the energy-integrated rate. \Fig{fig:diffrate} depicts the time-averaged differential rate (summed over isotopic composition) for a 100 GeV DM particle scattering off fluorine, sodium, iodine, argon, germanium, and xenon, for magnetic DM as a function of $\ER$. \Fig{fig:diffrate} includes both log-linear (left) and log-log (right) plots to show the different features of the spectra. If the differential rate were very steep, the integrated rate would be dominated by the differential rate at threshold, and thus have a similar annual modulation. As the differential rate flattens, an increasingly unweighted averaging occurs for all energies above threshold. The flattening of fluorine's differential rate occurs below PICO's $3.2$ keV threshold, and thus the pronounced features appearing in $\tmax$ of the differential rate should be strongly suppressed in the integrated rate.

\begin{figure*}[t!]
\centering
\includegraphics[width=.5\textwidth]{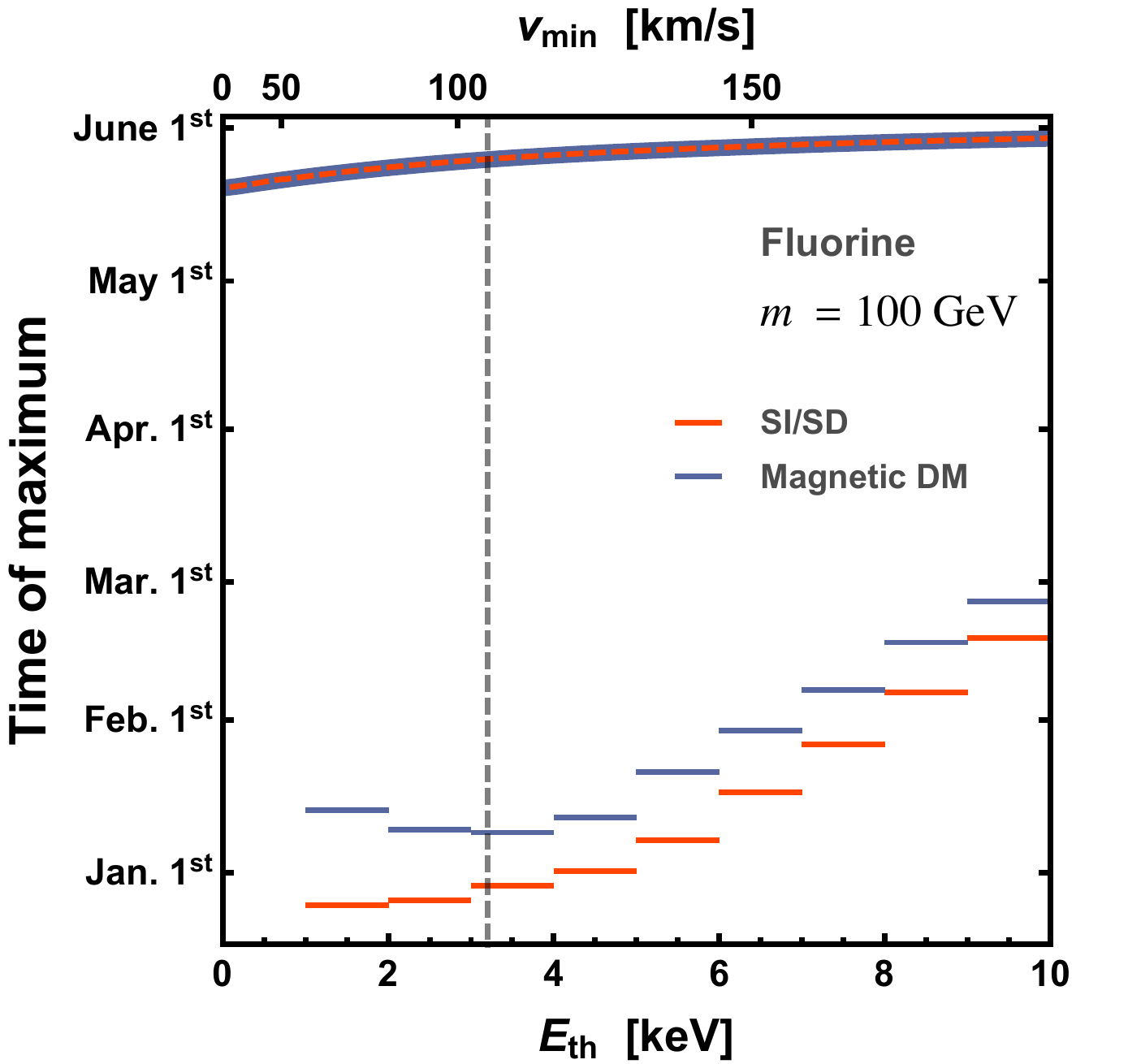}
\caption{\label{fig:TotRatePICO100} Time at which the energy-integrated rate is maximum as a function of threshold energy $E_\text{th}$ (the corresponding $\vmin$ value has been plotted on the upper horizontal axis), as observed by a fluorine detector for a $100$ GeV DM particle scattering elastically through a magnetic dipole (blue) and the standard SI/SD contact interaction (red). The solid and dashed lines depict the result of integrating the rate from a fixed threshold energy $E_\text{th}$, while horizontal bars show the result of binning data into $1$ keV bins. The efficiency function in \Eq{picoeffic} has been incorporated into all calculations. The vertical dashed line corresponds to PICO's $3.1$ keV lowest energy threshold.}
\end{figure*}

\Fig{fig:TotRatePICO100} depicts how PICO would realistically observe the time of maximum of the energy-integrated rate as a function of the threshold energy for a 100 GeV DM particle interacting through a magnetic dipole (solid blue line) or with the standard SI/SD contact interaction (dashed red line). As PICO does not provide an analytic form of their efficiency, we take the parametrization used by PICASSO,
\beq
\label{picoeffic}
\epsilon(\Ed) = 1 - e^{\alpha (1 - \Ed / E_\text{th})} \ ,
\eeq  
with $\alpha = 5$ for fluorine~\cite{Archambault:2012pm}. We also assume a perfect energy resolution, $G_{T}(\ER,\Ed) = \delta(\ER - \Ed)$. We have checked that the contribution from carbon is negligible for all energies so we consider only fluorine. \Fig{fig:TotRatePICO100} shows that the time of maximum of the rate as would be measured by PICO is nearly identical for the magnetic dipole interaction (dashed red line) and the standard SI/SD contact interactions (solid blue line), for all threshold energies we examined (larger than $0.1$ keV). To determine if the two interactions could be differentiated by binning the data, we also consider a fluorine-based experiment capable of measuring the rate in $1$ keV bins. For this hypothetical experiment we take the same efficiency function we used for PICO, and plot the result as horizontal bars in \Fig{fig:TotRatePICO100} for the magnetic dipole interaction (blue) and standard SI/SD contact interaction (red). The difference in the time of maximum of the energy-integrated rate for the two interactions in this hypothetical experiment ranges from $7$ days to $20$ days for threshold energies between $1$ and $10$ keV.

There are a number of reasons for the unique target-dependent features shown in \Fig{fig:Fmaxandmin} to be strongly suppressed when calculating the energy-integrated rate. First, the features in $\tmax$ for the magnetic dipole interaction differ the most from the standard SI/SD contact interactions in the $\vmin$ region where the $r_0$ and $r_1$ terms in \Eq{r0r1rate} cross over. For fluorine, this occurs at very small $\vmin$ values, $\vmin \lesssim 70$ km/s. The top axis of \Fig{fig:TotRatePICO100} shows that this region of $\vmin$ corresponds to very low energies, far below PICO's current threshold. Additionally, for elastic scattering $\ER \propto \vmin^2$, and since the integration of the differential rate is over $\ER$, the Jacobian's dependence on $\vmin$ must be included in the integrand when performing the integral in $\vmin$ instead. This additional factor increases the weight of the large $\vmin$ region in the integration where the modulation is target independent. Finally, as previously mentioned, the differential rate decreases rather slowly as a function of $\ER$, smearing the target-dependent features.

\begin{figure*}[t!]
\centering
\includegraphics[width=\textwidth, trim=0.2cm 1.6cm 0.1cm 1.5cm, clip=true]{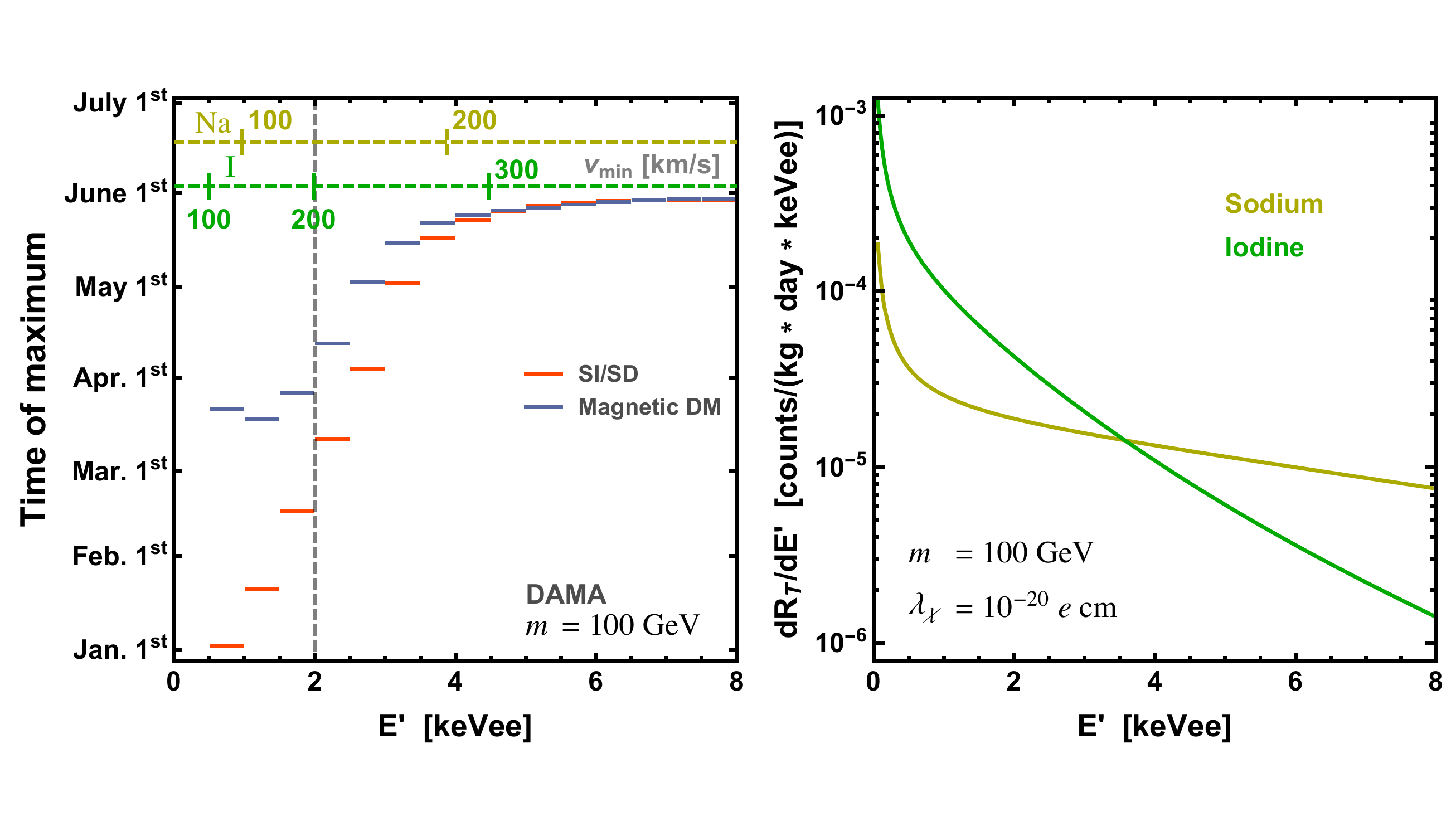}
\caption{\label{fig:DAMA} Left: $\tmax$ seen by DAMA for a $100$ GeV WIMP interacting through a magnetic dipole (blue) and the standard SI/SD contact interaction (red). Plotted with a vertical dashed line is the current DAMA low energy threshold. The horizontal dashed lines show the mapping of $\vmin$ onto $\Ed$ for sodium (yellow) and iodine (green) assuming quenching factors of $Q_\text{Na}=0.3$ and $Q_\text{I}=0.09$. Right: The time-averaged differential event rate for a $100$ GeV magnetic DM particle scattering off sodium (yellow) and iodine (green) as a function of detected energy.}
\end{figure*}

Let us see if other experiments could better preserve the target-dependent features.  Let us consider DAMA/LIBRA, henceforth referred to as DAMA (or any of the upcoming DAMA-like experiment as KIMS-NaI, ANAIS, DM-Ice17, and SABRE, see \eg\cite{Cooley:2014aya, Davis:2015vla} and references therein). DAMA is an interesting experiment to consider as both sodium and iodine have reasonably large nuclear magnetic moments and bin their data in small, $0.5$ keVee, intervals. In the left panel of \Fig{fig:DAMA} we plot the time of maximum of the DAMA binned rate as a function of $\Ed$ for both the magnetic dipole interaction (blue) and the standard SI/SD contact interaction (red), assuming elastic scattering with a $100$ GeV DM particle. Also depicted with a vertical dashed line is DAMA's current low energy threshold of $2$ keVee for the analysis of the modulated signal. The results for DAMA are calculated using quenching factors $Q_\text{Na}=0.3$ and $Q_\text{I}=0.09$, and a gaussian energy resolution function with standard deviation $0.448 \sqrt{\Ed} + 0.0091 \Ed$~\cite{Bernabei:2008yh}. The results for the two interactions are nearly indistinguishable above $4$ keVee, and only differ by about a month in the lowest observable energy bin. It is worth mentioning that DAMA will soon extend their low-energy threshold down to $1$ keVee which should result in a further observable difference between modulation arising from the standard SI/SD contact interactions and magnetic DM.

Like PICO, DAMA also sees a strong suppression in the target element dependent features of the modulation. The reason for the suppression in DAMA, however, is not primarily due to integrating over the differential rate, but rather due to the fact that DAMA has two non-negligible target elements. The independent contribution to the time-averaged differential rate from sodium (yellow) and iodine (green) as a function of detected energy is shown in the right panel of \Fig{fig:DAMA}. Since each element has a different $\vmin$ to $\Ed$ (average) mapping, and neither element dominates the differential rate in the $2$--$6$ keVee range, the target-dependent region of $\tmax$ for sodium is partially averaged with the target-independent region of iodine, leading to a large suppression of the target-dependent features. Furthermore, the small quenching factor of iodine pushes the most pronounced differences of the $\tmax$ curve below threshold. The horizontal dashed lines in the left panel of \Fig{fig:DAMA} show how the $\vmin$ values for sodium (yellow) and iodine (green) independently map to $\Ed$, in average.

\begin{figure*}[t!]
\centering
\includegraphics[width=0.49\textwidth]{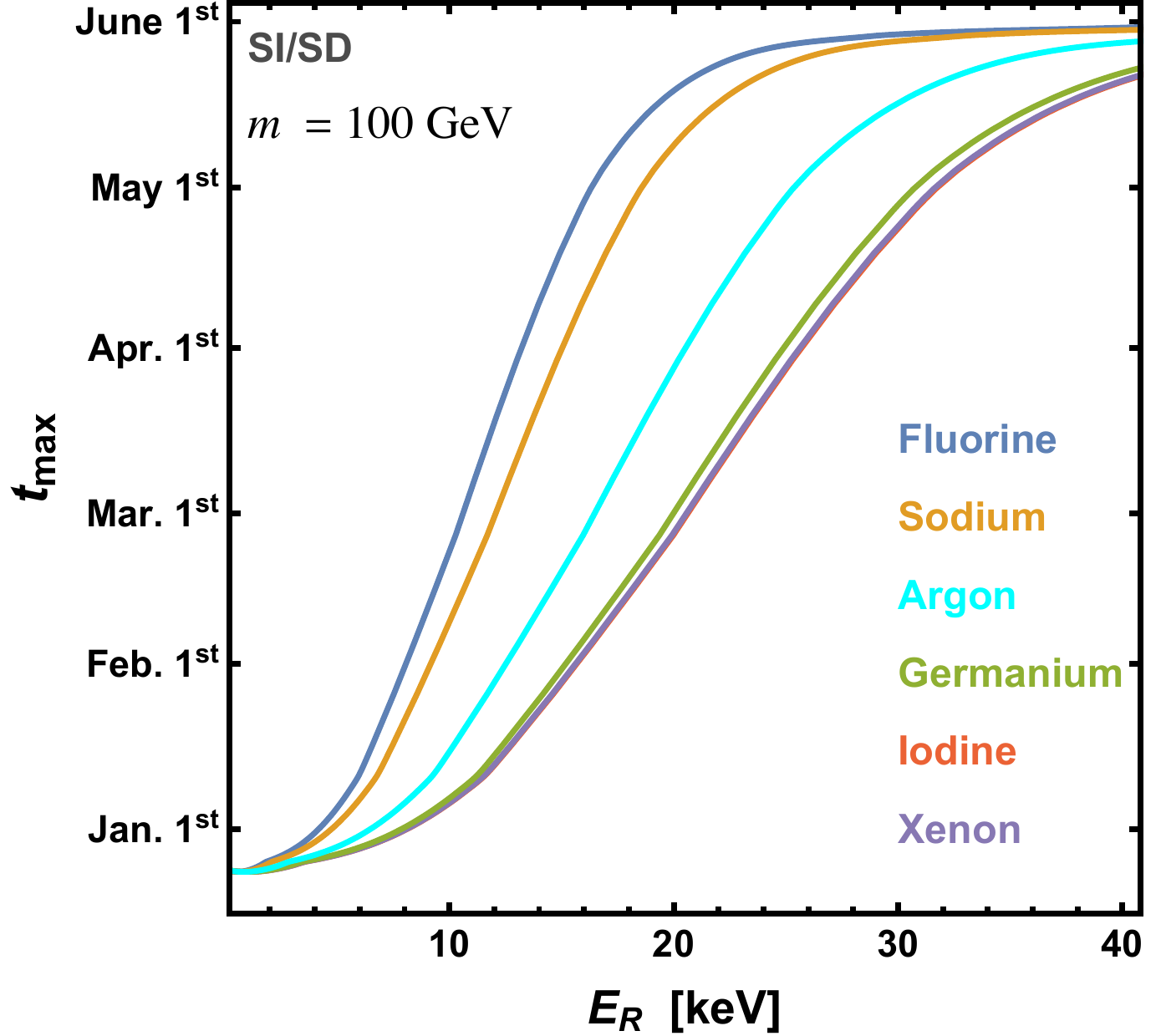}
\includegraphics[width=0.49\textwidth]{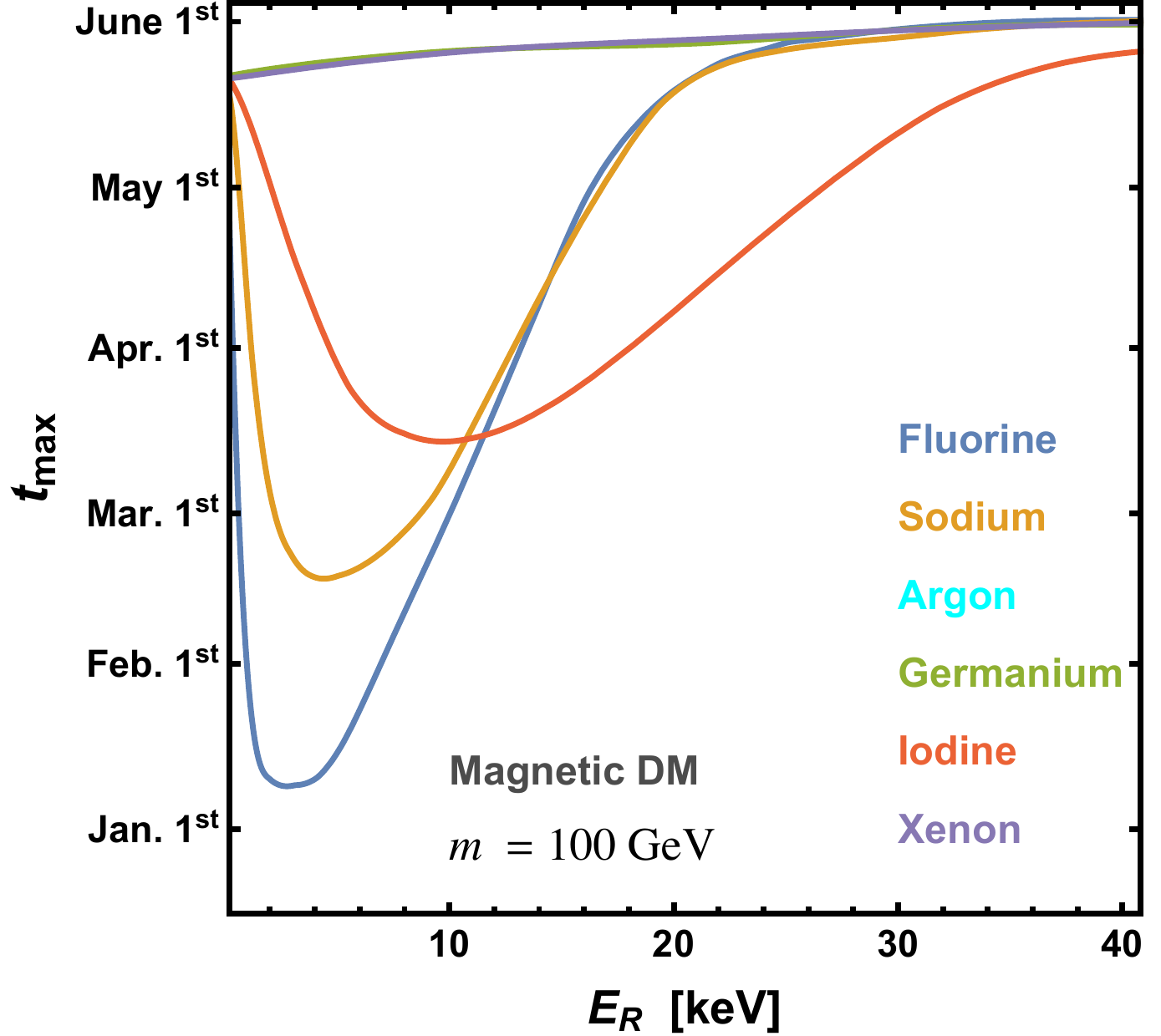}
\caption{\label{fig:MaxinER}  $\tmax$ for a $100$ GeV WIMP interacting with various elements through the standard SI/SD contact interaction (left) and a magnetic dipole (right) as a function of recoil energy. Note that the curves in the right panel for argon, germanium, and xenon all overlap and are nearly indistinguishable.}
\end{figure*}

Since experiments do not know the DM particle mass or the scattering kinematics a priori, it is nontrivial to obtain $\tmax$ as a function of $\vmin$ from the data. For this reason, and because $\tmax$ as a function of $\ER$ is necessarily known to be target element dependent, it is logical to ask how $\tmax$ for magnetic DM differs from $\tmax$ for the standard SI/SD contact interactions as functions of $\ER$. This comparison is made in \Fig{fig:MaxinER}, where the left panel shows $\tmax$ for SI/SD interactions while the right panel shows $\tmax$ for magnetic DM, both as functions of $\ER$. In both cases we assume a $100$ GeV DM particle scattering elastically with various target elements (note that the curves for argon, germanium, and xenon in the right panel overlap almost entirely).

For the standard contact interaction with only $r_0$ in the rate (see Eqs.~\eqref{r0r1rate} and \eqref{r0r1}), as the SI/SD interaction, the differences in the curves is determined solely by the mass of the target nuclide. The largest difference in $\tmax$ therefore occurs between fluorine and xenon and is around three months for recoil energies between $15$ and $20$ keV. While this is a rather large discrepancy, the shape of the $\tmax$ curves for the standard SI/SD interactions are all stretched and compressed images of each other. In fact, all curves are obtained from the curve for $\eta_0$ in \Fig{fig:SHM} with the $\ER$-$\vmin$ relation for elastic scattering in \Eq{vmin}, which of course only differ in each case for the choice of $m_T$. In this sense, $\tmax$ and other observables associated with the modulation are not truly target dependent for interactions with only $r_0$ in the rate. The same cannot be said for magnetic DM. The right panel of \Fig{fig:MaxinER} shows that the difference between various $\tmax$ curves is more pronounced than when the standard interactions are considered, and furthermore, the curves have a more individualized shapes. The only exception are the curves for germanium, argon, and xenon, which completely overlap for a $100$ GeV DM particle, a consequence of having a small or zero (for argon) average nuclear magnetic moment.

\subsection{Inelastic scattering}

Prior to this point we have only considered DM-nuclei elastic scattering. It has been shown that inelastic scattering, which can occur when there exist at least two DM particles with nearly degenerate masses $m$ and $m + \delta$ with $\delta \ll m$, has the potential to significantly alter the scattering kinematics and the observed annual modulation~\cite{TuckerSmith:2001hy,Graham:2010ca}.

Inelastic endothermic scattering occurs when the light DM state scatters into the heavy DM state, $\delta > 0$. Since this process requires additional energy, only DM particles traveling at speeds greater than or equal to $v_\delta^{T} \equiv \sqrt{2 \delta / \mu_T}$ can scatter off a particular target $T$. If GF is the sole source of anisotropy, target-dependent modulation can only occur when speeds of about $200$ km/s are probed. This implies that for a fixed DM mass, there exists a maximum mass splitting $\delta_\text{max}$ for which target-dependent modulation can occur. For a $100$ GeV DM particle scattering off fluorine, sodium, and iodine, this corresponds to values of $\delta_\text{max} \approx 3.3$ keV, $4$ keV, and $12$ keV, respectively. These values of $\delta$ are quite small with respect to the typical momentum transfer in the interaction, and thus we expect the scattering kinematics to be almost elastic. Without an additional form of anisotropy, endothermic scattering is therefore ineffective in probing values of $\vmin$ which can lead to target-dependent modulation.

\begin{figure}[t!]
\centering
\includegraphics[width=\textwidth, trim=0cm 1.3cm 0.1cm 1.5cm, clip=true]{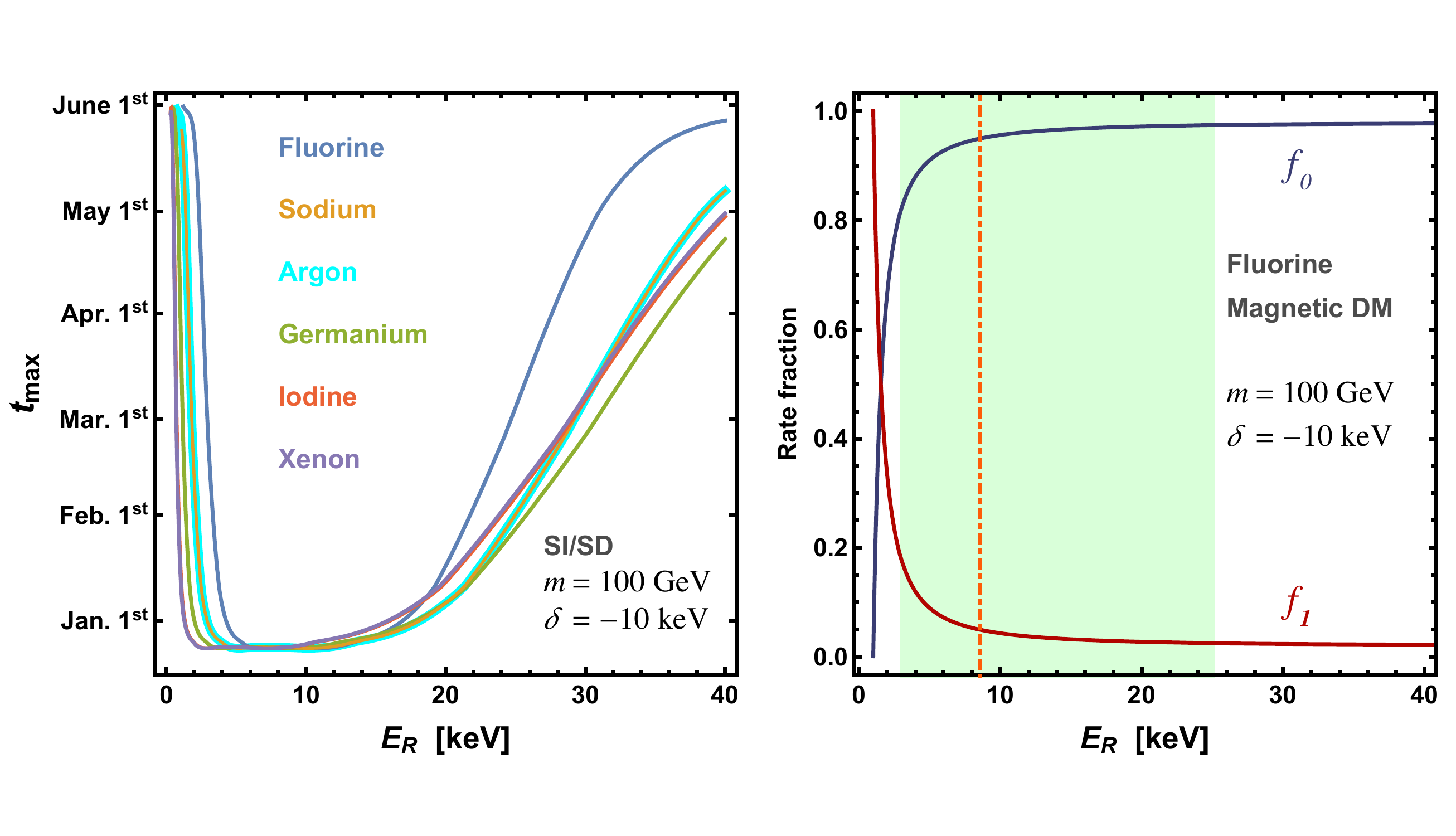}
\caption{\label{fig:exothermic} Left: $\tmax$ for the exothermic scattering with various elements assuming the standard SI/SD contact interactions, a DM mass of $100$ GeV, and a mass splitting $\delta = -10$ keV, as a function of $\ER$. Right: Rate fractions $f_0$ (blue) and $f_1$ (red) for magnetic DM (as defined in Sec.~\ref{mdm}) for $100$ GeV DM scattering off fluorine, assuming $\delta = -10$ keV. The shaded green region highlights recoil energies corresponding to values of $\vmin < 200$ km/s, and the dot-dashed orange line depicts the $\ER$ value corresponding to $\vmin = 0$ km/s. }
\end{figure}

Inelastic exothermic scattering, occurring when the heavier DM particle down-scatters into the lighter DM state ($\delta < 0$), can be potentially more interesting for target-dependent modulation. To illustrate how exothermic scattering can alter the observed modulation, we plot in the left panel of \Fig{fig:exothermic} $\tmax$ for DM interacting with various elements through the standard SI/SD contact interaction, assuming $m = 100$ GeV and $\delta = -10$ keV, as a function or $\ER$. This result is obtained by mapping the $\taumax(\vmin)$ line corresponding to $\eta_0$ shown in \Fig{fig:SHM} onto $\ER$ by using the $\ER$-$\vmin$ relation for inelastic scattering,
\beq
\label{scatinelastic}
\vmin (\ER) = \frac{1}{\sqrt{2 m_T \ER}} \left| \frac{m_T \ER}{\mu_T} + \delta \right|
\eeq
(remember that for the SI/SD interaction $\tmax$ coincides with $\taumax$).

We have chosen not to plot $\tmax$ for magnetic exothermic dark matter because, for all elements considered, the results mirror what would be expected should the differential cross section either be independent of velocity, or proportional to $v^{-2}$. That is to say for a given element, only the term proportional to $\eta_0$ or the term proportional to $\eta_1$ is relevant, never both. To understand why this is the case, it is necessary to first consider the differential cross section~\cite{Chang:2010en}:
\begin{multline}
\label{diffsigmamagneticinelastic}
\frac{\ud \sigma_T}{\ud \ER}(\ER, v) =
\alpha \lambda_\chi^2 \left\{ \frac{Z_T^2}{\ER} \left[ 1 - \frac{\ER}{v^2} \left( \frac{1}{2 m_T} - \frac{1}{m^2} \right) \right. \right. \\  \left. \left. -\frac{\delta}{v^2} \left( \frac{1}{\mu_T} + \frac{\delta}{2 m_T \ER}\right) \right] F_{\text{SI}, T}^2(\ER) +  \frac{\hat\lambda_T^2}{v^2} \frac{m_T}{m_p^2} \left( \frac{S_T + 1}{3 S_T} \right) F_{\text{M}, T}^2(\ER) \right\} .
\end{multline}
There are two additional terms with respect to the elastic case in \Eq{diffsigmamagnetic}, both contributing to the charge-dipole term for inelastic magnetic dark matter, one of which is proportional to $\ER^{-1}$ and the other to $\ER^{-2}$. Both of these terms are contained within $f_0$ (see Sec.~\ref{mdm}), and since the target dependence relies on the interplay between $f_0$ and $f_1$, it is important to understand how these two new terms contribute to the relative rate fractions.

In Sec.~\ref{mdm}, we showed that for elastic scattering $f_1$ is always the dominant contribution to the rate at low $\vmin$. This is a consequence of having a term proportional to $\vmin^{-2} \propto \ER^{-1}$. For inelastic magnetic DM, $f_0$ now has a term proportional to $\ER^{-2}$, thus at very low energies $r_0$ is always the dominant contribution to the rate. This might be avoided, however, because there may exist a lower limit on $\ER$ which depends on $\vesc$, and this may be above the region where $\ER^{-2}$ is the dominant factor (see Fig.~1 of~\cite{DelNobile:2015lxa}). At large energies, both of the new terms will be suppressed, and as for elastic scattering, the rate should be controlled by the term containing the magnetic form factor, $r_0$ (assuming the target element has a non-negligible nuclear magnetic moment). Whether $r_0$ or $r_1$ dominates the rate at intermediate energies depends strongly on the target element, the DM mass, and $\delta$.

To illustrate how these variables affect the potential appearance of target-dependent modulation, we plot in the right panel of \Fig{fig:exothermic} the rate fraction for magnetic exothermic DM scattering off fluorine, assuming $m = 100$ GeV and $\delta = -10$ keV. The blue and red lines show the terms proportional to $f_0$ and $f_1$, respectively. The green region highlights values of $\ER$ where target-dependent modulation could potentially be observed (\ie $\vmin \lesssim 200$ km/s, assuming GF is the sole source of anisotropy), and the dot-dashed orange line depicts the energy corresponding to $\vmin = 0$ km/s. To compute the rate we again use the form factors provided in~\cite{Fitzpatrick:2012ix, Fitzpatrick:2012ib}. While these only apply to elastic scattering, \cite{Barello:2014uda} showed that they can be adapted to inelastic scattering by properly taking into account the modification to $\bol{v}^\perp = \bol{v} + \bol{q} / 2 \mu_N$, the component of $\bol{v}$ orthogonal to the momentum transfer $\bol{q}$, due to inelastic kinematics ($\mu_N$ being here the DM-nucleon reduced mass). Therefore one simply needs to replace the variable $\bol{v}^\perp$ in the form factors of~\cite{Fitzpatrick:2012ix, Fitzpatrick:2012ib} with the true orthogonal component of the DM velocity for inelastic scattering, $\bol{v}^\perp_\text{inel} = \bol{v}^\perp + \delta \bol{q} / |\bol{q}|^2$.

Two comments are in order. We previously stated that $f_0$ should be the dominant term at low values of $\ER$ due to the $\ER^{-2}$ term in the differential cross section. While this may not appear to be the case in \Fig{fig:exothermic}, this is simply because we have not plotted the low $\ER$ regime, as it is not relevant for target-dependent modulation (low $\ER$ corresponds to large WIMP velocities where GF is unimportant). Next, for the current choice of parameters, $f_0$ is the only relevant term in the $\ER$ range where the effect GF is important, and thus the $\tmax$ curve is identical to the fluorine curve shown in the left panel of \Fig{fig:exothermic}. We stress that the unique target-dependent features seen in the $\tmax$ and $\tmin$ curves of Figs.~\ref{fig:Fmaxandmin}--\ref{fig:Xemaxandmin} only arise if both $f_0$ and $f_1$ contribute in a non-negligible way within the region capable of probing low DM speeds.

It is interesting to see how changing $m$, $\delta$, and the target element alter the results of \Fig{fig:exothermic}. Changing the DM particle mass results in two distinct effects. Contrary to elastic scattering, lower values of $m$ increase the importance of $f_0$ relative to $f_1$ at fixed $\ER$, and thus the point at which $f_0$ becomes dominant relative to $f_1$ shifts to lower values of $\ER$. The second and more import effect arises from changing the value of $m$ in \Eq{scatinelastic}, which causes the $\ER$ range where the effect of GF is relevant in the right panel of \Fig{fig:exothermic} to shift. Using \Eq{scatinelastic}, one can see that decreasing the DM mass shifts the influence of GF to lower values of $\ER$. We have checked that for $\delta \geqslant 10$ keV, lowering the DM particle mass to $10$ GeV does not bring the point at which $f_0$ and $f_1$ cross into the region where target dependent modulation could occur.     

Increasing the magnitude of $\delta$ (\ie making $\delta$ more negative) also has two effects. First, it shifts the point at which $f_0$ and $f_1$ cross to higher values of $\ER$. This effect is completely negligible, however, when compared with how this change in $\delta$ shifts the $\ER$ range where the effect of GF is relevant (see \Eq{scatinelastic}). 

The negligible nuclear magnetic moments of germanium, xenon, and argon lead to a complete dominance of $f_1$ over $f_0$ for essentially all values of $\ER$, regardless of the DM mass and $\delta$. This implies that inelastic magnetic DM scattering with these elements will always lead to an observation of $\tmax$ between late May and early June, and the annual modulation will be consistent with inelastic scattering through differential cross sections that are independent of velocity. For iodine and sodium we have checked that the crossover from $f_1$ to $f_0$ as the dominant contribution to the rate, either always occurs far below threshold, or does not occur in the region where target-dependent modulation would arise. Identifying this type of scattering would then necessitate at least one experiment employing germanium, xenon, or argon, and another experiment employing fluorine, sodium, or iodine, to observe the annual modulation.

\subsection{Identification of non-factorizable cross sections \label{identification}}

The target-dependent effects described thus far have relied on two assumptions: experiments probe anisotropy in the DM halo and velocity and target dependence cannot be factored in the DM-nucleus differential scattering cross section. The question remains how a differential cross section of this form could be identified. A single experiment can never uniquely determine the underlying particle physics and astrophysics; it is only possible for a single experiment to say that their findings are consistent with some set of assumptions on the distribution of DM, the DM mass, a particular DM-nucleus interaction, etc. The most model-independent information is likely to come from a comparison of the outcomes of different experiments. We believe the most effective way to confirm the existence of a DM-nucleus cross section with a non-factorizable target and velocity dependence is to show that there exists no $\ER$-$\vmin$ relation capable of mapping observables associated with the modulation of the rate from experiments employing different target elements onto a unique function of $\vmin$. We emphasize however that finding unique functions of $\vmin$ capable of reconciling the results of multiple experiments does not preclude the existence of non-factorizable differential cross sections. In the case of inelastic magnetic DM, elements with small average nuclear magnetic moments, \eg germanium, xenon, argon, and carbon, will all yield similar results because the differential cross section is dominated by a single term, at least for the $\vmin$ region where the local DM distribution is made anisotropic by GF.

\section{Conclusions}
\label{conclusion}
It is typically assumed that observables associated with the annual modulation of the rate in direct detection experiments, when expressed as functions of $\vmin$ (the minimum DM speed necessary to impart a given recoil energy to a target nucleus), are unique target-independent functions. We have shown that this is not necessarily the case, and in fact the existence of a DM-nucleus differential cross section with a non-factorizable target and velocity dependence naturally leads to target-dependent modulation. The identification of this type of differential cross section is not straightforward and must be done through a process of elimination. In the event that multiple experiments with putative signals cannot find an $\ER$-$\vmin$ relation that can reconcile the differences between the observed modulations, one may then infer the potential existence of a non-factorizable differential cross section. We emphasize, however, that the reverse is not true. That is to say, finding an $\ER$-$\vmin$ relation that maps observables associated with the modulation from multiple experiments onto unique $\vmin$-dependent functions does not necessarily ensure that the modulation is target independent.

As a specific example, we have shown how $\tmax$ ($\tmin$), the time of maximum (minimum) of the differential rate, depends on the target nuclide for magnetic dipole DM elastically scattering with fluorine, germanium, iodine, sodium, argon, and xenon. We have also discussed how the annual modulation would appear should DM scatter inelastically with these elements. In our calculations we assume the SHM and included the effect of GF. We have shown that in an idealized experiment, the observed difference in $\tmax$ for DM scattering off fluorine and xenon at a fixed value of $\vmin$ could differ by as much as four months for DM masses above $50$ GeV, however, accounting for the limitations of a realistic detector and integrating the differential rate can significantly suppress these differences. The plausible presence of DM substructure or forms of anisotropy other than GF could nevertheless enhance the target dependence of the modulation.

\section{Acknowledgments} 
E.D.N.~and G.G.~acknowledge partial support from the Department of Energy under Award Number DE-SC0009937. E.D.N.~was also supported in part by the MIUR-FIRB Grant RBFR12H1MW.

\bibliographystyle{JHEP}
\bibliography{biblio}

\end{document}